\documentclass[12pt]{article}  
\usepackage{a41}
\usepackage{cite}
\usepackage{amsmath}
\usepackage{amssymb}
\usepackage{color}  

\usepackage[rflt]{floatflt}              
\usepackage{float}
\usepackage{slashed}

\def\b0{\beta_0}

\usepackage{array}

\newcommand{\HA}{{\rm H}}

\usepackage[english]{babel}
\usepackage[latin1]{inputenc}
\usepackage[T1]{fontenc}
\usepackage{ae}

\usepackage{url}

\usepackage{amsmath, amsthm, amssymb}
\newtheorem{thm}{Theorem}[section]

\newtheorem{definition}[thm]{Definition}

\newcommand{\gsim}{\raisebox{-0.07cm}
{$\, \stackrel{>}{{\scriptstyle\sim}}\, $}}
 \newcommand{\GeV}{\mathrm{GeV}}

\newcommand{\Li}{{\rm Li}}

\newcommand{\Mvec}{{\rm\bf M}}

\newcommand{\ep}{\varepsilon}

\usepackage{rotating}

\usepackage{graphicx}

\newcounter{mmacnt}
\def\restartmma{\setcounter{mmacnt}{0}}
\restartmma \catcode`|=\active
\def|#1|{\mathrm{#1}}
\catcode`|=12
\newenvironment{mma}{
 \par\smallskip
 \catcode`|=\active
 \parskip=0pt\parindent=0pt 
 \small
 \def\In##1\\{%
\def\linebreak{\hfill\break\null\qquad}%
\refstepcounter{mmacnt}
\hangindent=2.5em\hangafter=0
\leavevmode
\llap{\tiny\sffamily n[\arabic{mmacnt}]:=\kern.5em}%
\mathversion{bold}\footnotesize$\displaystyle##1$\normalsize
\mathversion{normal}\par
 }%
 \def\Print##1\\{%
\def\linebreak{\hfill\break}%
\hangindent=2.5em\hangafter=0
\leavevmode ##1\par}%
 \def\Out##1\\{%
\def\linebreak{$\hfill\break\null\hfill$}%
\kern\abovedisplayskip\par
\hangindent=2.5em\hangafter=0
\leavevmode
\llap{\tiny\sffamily Out[\arabic{mmacnt}]=\kern.5em}
\footnotesize$\displaystyle##1$\normalsize\hfill\null\par
\kern\belowdisplayskip
 }%
 \def\Warning##1##2\\{%
\def\linebreak{\hfill\break}%
\hangindent=2.5em\hangafter=0
\leavevmode
{\scriptsize##1 : ##2}\par}%
}{%
 \par\smallskip
}

\usepackage{color}

\newenvironment{fshaded}{%
\MakeFramed {\FrameRestore}
}%
{\endMakeFramed}

\def\b0{\beta_0}

\def\Gp0{{\Gamma^{'}_0}}
\def\Gp1{{\Gamma^{'}_1}}
\def\Gp2{{\Gamma^{'}_2}}

\allowdisplaybreaks[4]

\begin{document}
\setlength{\baselineskip}{0.515cm}

\sloppy
\thispagestyle{empty}
\begin{flushleft}
DESY 19--060
\\
DO--TH 19/04\\
\end{flushleft}

\mbox{}
\vspace*{\fill}
\begin{center}

{\LARGE\bf The Polarized Two-Loop Massive}

\vspace*{2mm}
{\LARGE\bf Pure Singlet Wilson Coefficient}

\vspace*{2mm}
{\LARGE\bf for Deep\hspace*{0.1mm}-\hspace*{-0.5mm}Inelastic Scattering}

\vspace{3cm}
\large
{\large 
J.~Bl\"umlein$^a$, 
C.~Raab$^b$,
and
K.~Sch\"onwald$^{a}$
}

\vspace{1.cm}
\normalsize
{\it   $^a$~Deutsches Elektronen--Synchrotron, DESY,}\\
{\it   Platanenallee 6, D--15738 Zeuthen, Germany}
              
\vspace*{2mm}
{\it  $^b$~Johannes Kepler Universit\"at Linz,}\\
{\it Altenberger Stra\ss{}e 69, A--4040 Linz, Austria}


\end{center}
\normalsize
\vspace{\fill}
\begin{abstract}
\noindent
We calculate the polarized massive two--loop  pure singlet Wilson coefficient contributing to the structure 
functions $g_1(x,Q^2)$ analytically in the whole kinematic region. The Wilson coefficient contains Kummer--elliptic 
integrals. We derive the representation in the asymptotic region $Q^2 \gg m^2$, retaining power corrections, and 
in the threshold region. The massless Wilson coefficient is recalculated. The corresponding twist--2 corrections 
to the structure function $g_2(x,Q^2)$ are obtained by the Wandzura--Wilczek relation. Numerical results are presented.
\end{abstract}

\vspace*{\fill}
\noindent
\newpage 
\section{Introduction}
\label{sec:1}

\vspace*{1mm}
\noindent
Analytic expressions in perturbative Quantum Chromodynamics, as for partonic sub--system scattering cross 
sections and Wilson coefficients, provide an excellent basis for both numerical studies, fits to precision data 
and provide analytic insight into the structure of Feynman integrals. Analytic results also allow to derive
important limiting cases and make it easier to incorporate resummations in specific kinematic regions. With a 
growing number of scales and loops for the respective processes it becomes more difficult to obtain analytic 
results, although there has been significant progress in analytic integration methods recently; for a survey 
cf.~\cite{Blumlein:2018cms}. The precise knowledge of these corrections is of importance to measure the 
polarized parton densities in high energy collisions and to  determine, related to it, the 
strong coupling constant $\alpha_s(M_Z)$ and the heavy quark masses, cf.~\cite{Boer:2011fh}.

The first two--loop QCD heavy flavor corrections to the polarized structure function $g_1(x,Q^2)$ have been
calculated in \cite{Buza:1996xr} in the asymptotic region $Q^2 \gg m^2$, where $Q^2$ denotes the virtuality of 
the exchanged photon and $m$ the mass of the heavy quark. The asymptotic two--loop QCD corrections have been 
recalculated in \cite{Bierenbaum:2007zz,POL18}. In \cite{Buza:1996xr} the region of low values of $Q^2$ has 
been modeled by an ansatz. The leading threshold resummation for the gluonic contributions has been studied 
in \cite{Eynck:2000gz}. The complete two--loop polarized heavy flavor Wilson coefficient in the non--singlet case 
has been calculated analytically in the tagged flavor case in \cite{Buza:1996xr} and for the complete 
contribution to the structure function $g_1(x,Q^2)$ in \cite{Blumlein:2016xcy}, also completing former work on
the polarized Bjorken sum rule in \cite{Blumlein:1998sh}. Numerical results for the polarized two--loop heavy 
flavor case have been given in \cite{Hekhorn:2018ywm} recently. Finally, in the non--singlet case the asymptotic
contributions have been calculated to three--loop order analytically in \cite{Behring:2015zaa,Ablinger:2014vwa}.

In the present paper we follow Ref.~\cite{Blumlein:2019qze} in the unpolarized case and calculate the polarized 
pure singlet two--loop 
heavy flavor corrections for the structure function $g_1(x,Q^2)$ in the whole kinematic range analytically. We 
also compute the corresponding massless contributions, which have first been calculated in 
\cite{Zijlstra:1993sh} and later in \cite{Vogt:2008yw}.
Since the calculations are carried out using dimensional regularization in $D = 4 + \ep$ dimensions one may 
work in the Larin--scheme \cite{Larin:1993tq}\footnote{For other $\gamma_5$ schemes see Refs.~\cite{HVBM}.}  
and perform a finite renormalization to the {\sf M}--scheme \cite{Matiounine:1998re,Moch:2014sna} afterwards. 
We derive 
both the result in the asymptotic range $Q^2 \gg m^2$, see also Refs.~\cite{Buza:1996xr,POL18}, and in the 
threshold 
region. Numerical results are presented. Various technical aspects of the present calculation can be found in 
Ref.~\cite{Blumlein:2019qze} already.

The paper is organized as follows. In Section~\ref{sec:2} we summarize basic relations for the polarized 
deep--inelastic scattering cross section. In Section~\ref{sec:3} the result for the massless pure singlet 
Wilson coefficient $C_{g_1}^{(2),\rm PS}$ is presented. The recalculation of the massless Wilson coefficient
is necessary, since in Ref.~\cite{Zijlstra:1993sh} different schemes have been used in part. The corresponding 
massive Wilson coefficient is calculated in Section~\ref{sec:4}. The corresponding results for the twist--2 
contributions to the structure function $g_2(x,Q^2)$ can be obtained by using the Wandzura--Wilczek relation 
\cite{Wandzura:1977qf}, as has been shown for the massless quarkonic \cite{G2A,G2B}  and gluonic \cite{Blumlein:2003wk} 
cases, for diffractive scattering \cite{Blumlein:2002fw}, non--forward scattering \cite{Blumlein:2000cx}, and the 
target mass corrections \cite{BT,TM}. Limiting cases are studied in Section~\ref{sec:5} and numerical results 
are presented in Section~\ref{sec:6}. The conclusions are given in Section~\ref{sec:7}. Some Mellin convolutions 
appearing due to renormalization are listed in the Appendix.

\vspace*{-.3cm}
\section{The Deep-inelastic Scattering Cross Section}
\label{sec:2}

\vspace*{1mm}
\noindent
The scattering cross sections for deep--inelastic charged lepton scattering of polarized nucleons 

\newpage
\noindent
are obtained polarizing the incoming lepton longitudinally and the target nucleon either longitudinally or 
transversally, resulting into the spin 4-vectors $S_L$ and $S_T$,
\begin{eqnarray}
S_L &=& (0, 0, 0; M) \\
S_T &=& M(0, \cos(\beta), \sin(\beta); 0)
\end{eqnarray}
in the nucleon rest frame. One has $S_L.p = S_T.p =0$, with $p$ the nucleon 4-momentum.
The scattering cross sections are given by, cf.~e.g.~\cite{BT,Lampe:1998eu},
\begin{eqnarray}
\frac{d^2 \sigma(\lambda, \pm S_L)}{dx dy} &=& \pm 2 \pi S \left[-2 \lambda y \left(2 - y - \frac{2xy M^2}{S}\right)
x g_1(x,Q^2)  + 8 \lambda \frac{y x M^2}{S} x g_2(x,Q^2) \right] 
\\
\frac{d^3 \sigma(\lambda, \pm S_T)}{dx dy d\phi} &=& \pm S \frac{\alpha^2}{Q^4} 2 \sqrt{\frac{M^2}{S}} 
\sqrt{xy \left[1-y - \frac{xyM^2}{S}\right]} \cos(\beta-\phi) 
\nonumber\\ && \times
\left[-2 \lambda y x g_1(x,Q^2) - 4 \lambda x 
g_2(x,Q^2)\right]
\end{eqnarray}
for pure virtual photon exchange. Here $S$ denotes the cms energy of the process, $M$ is the nucleon mass,
$\lambda$ the degree of lepton polarization, $\alpha = e^2/(4\pi)$ is the fine structure constant, 
$Q^2 = -q^2$ denotes the photon virtuality and $x = Q^2/(Sy), y = l.q/p.q$ are the Bjorken variables with 
$l$ the incoming charged lepton and proton momenta, $S = (p+l)^2$ and $\phi$ is the 
azimuthal angle of the final state lepton, which can 
be integrated over in the case of longitudinal polarization.

In the following we will present a series of relations in Mellin--$N$ space for convenience. The respective 
quantities
in momentum-fraction $z$--space are related to those in Mellin--space by the transformation
\begin{eqnarray}
\label{eq:MEL}
\Mvec\left[f(z)\right](N) = \int_0^1 dz z^{N-1} f(z). 
\end{eqnarray}
The structure function $g_1(N,Q^2)$ is given in the twist--two approximation using the factorization 
theorems \cite{FACT} by
\begin{eqnarray}
g_1(N,Q^2) &=& \frac{1}{2} \Biggl[
\frac{1}{N_F} \sum_{k=0}^{N_F} e_k^2 \left\{\Sigma(N,\mu_F^2) C_q^{\rm PS}\left(N,\frac{Q^2}{\mu_F^2}\right) 
+
G(N,\mu_F^2) C_g^{\rm S}\left(N,\frac{Q^2}{\mu_F^2}\right) \right\} 
\nonumber\\ && 
+ \Delta\left(N,\mu_F^2\right) C_q^{\rm NS}\left(N,\frac{Q^2}{\mu_F^2}\right) \Biggr].
\end{eqnarray}
Here 
\begin{eqnarray}
\Sigma(N) = \sum_{k=1}^{N_F} \left[ q(N) + \bar{q}(N) \right]
\end{eqnarray}
denotes the singlet distribution, $G(N)$ the gluon distribution, $\Delta(N)$ the flavor non--singlet distribution
\begin{eqnarray}
\Delta(N) = \sum_{i=1}^{N_F} \left[e_i^2 - \frac{1}{N_F} \sum_{k=1}^{N_F} e_k^2\right] \left[q_i(N) + \bar{q}_i(N) 
\right].
\end{eqnarray}
$e_k$ labels the electric charge of the $k$th light quark.

The Mellin transform of the structure function $g_2$ is related to that of $g_1$ by 
\begin{eqnarray}
g_2(N,Q^2) = - \frac{N-1}{N} g_1(N,Q^2)
\end{eqnarray}
or
\begin{eqnarray}
g_2(x,Q^2) = - g_1(x,Q^2) + \int_0^1 \frac{dy}{y} g_1(y,Q^2).
\end{eqnarray}
Note that in the massive pure singlet case the support of both structure functions in limited by $0 < x < 
1/(1+4 m^2/Q^2)$ due to the production of two heavy quarks.

The different steps in the renormalization and factorization of the polarized massless Wilson coefficients 
have been described in \cite{Zijlstra:1992qd,Zijlstra:1993sh} and for the massive Wilson coefficients in
\cite{POL18} using the Larin--scheme. In the present case the finite renormalization moving to the 
{\sf M}--scheme only affects the massless or massive Wilson coefficient by adding the term $-z_{\rm PS}^{(2)}$, 
Eq.~(\ref{eq:ZPS2}).
\section{The Massless Wilson Coefficient}
\label{sec:3}

\begin{figure}[H]
      \centering  
      \includegraphics[width=0.5\textwidth]{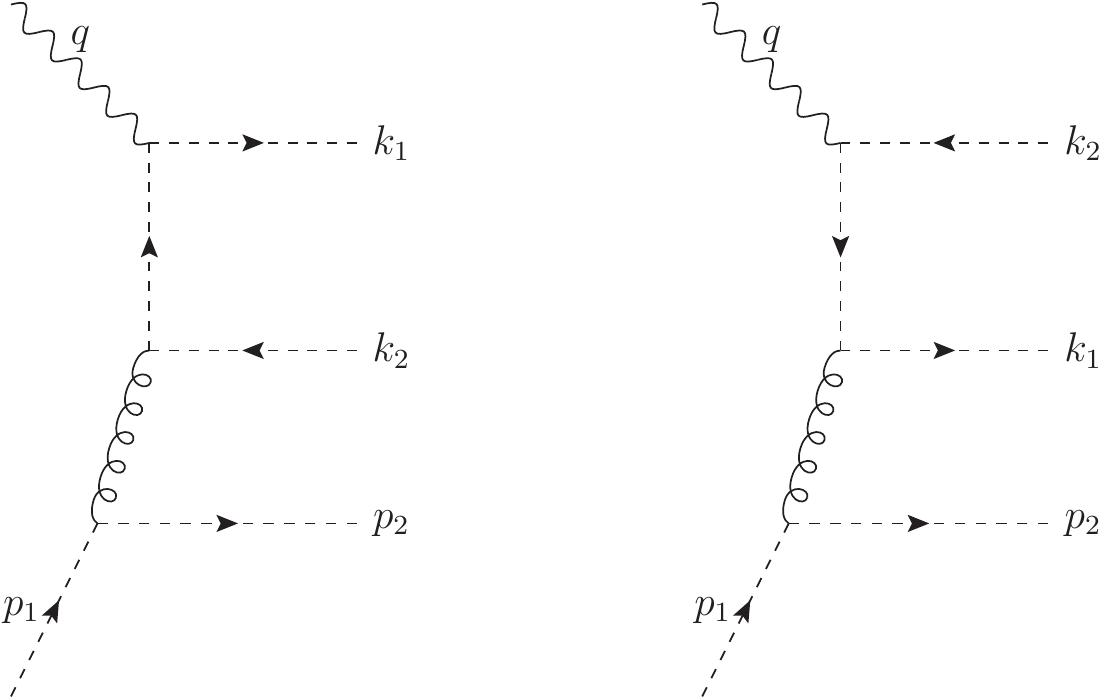}
      \caption{\small \sf The Feynman diagrams for the massless and massive two--loop pure singlet Wilson 
coefficient.}
      \label{fig:FIG2}
\end{figure}

\vspace*{1mm}
\noindent
The Feynman diagrams contributing to the polarized massless two--loop Wilson coefficient are shown in 
Figure~\ref{fig:FIG2}.
\noindent
forming the amplitude squared $G_{\mu\nu}$, where the indices refer to the coupling of the virtual photon. 
Here all quark lines are massless. 
The massless resp. massive Wilson coefficients are obtained following Ref.~\cite{Zijlstra:1993sh}, 
Eqs.~(3.7--3.18).  The corresponding phase space integrals have been given in Ref.~\cite{Blumlein:2019qze}. 
We apply the Larin--scheme \cite{Larin:1993tq} in which the contraction of the free indices of the 
two appearing Levi--Civita tensors have to be performed in $D$ dimensions.
 
\begin{figure}[H]
      \centering  
      \includegraphics[width=0.5\textwidth]{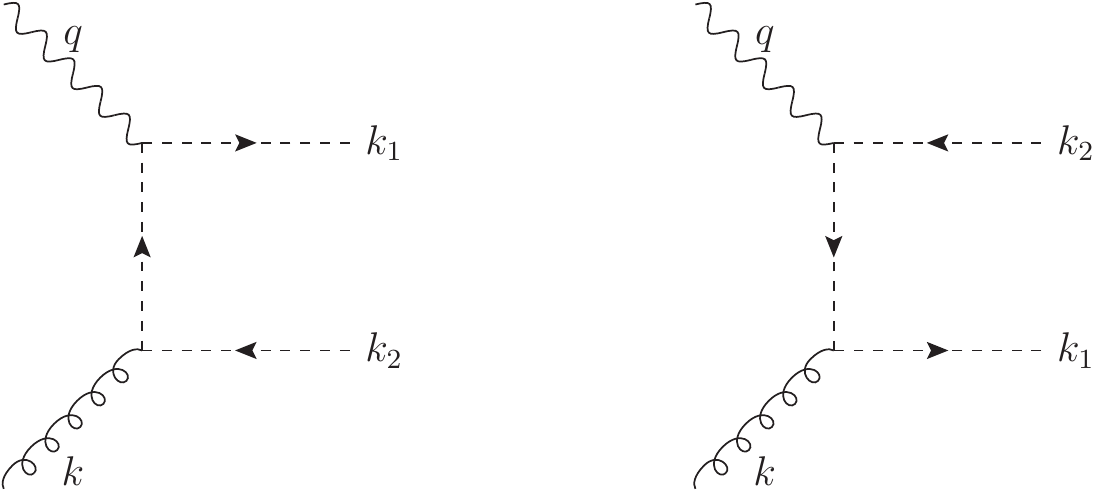}
      \caption{\small \sf The one--loop Feynman diagrams.}
      \label{fig:FIG1}
\end{figure}

The unrenormalized two--loop massless pure singlet Wilson coefficient reads  in Mellin--$N$ space
\begin{eqnarray}
\hspace*{-3mm}
\hat{\hat{C}}_{g_1}^{(2),\rm PS} &=& \hat{a}_s^2 S_\ep^2 \left(\frac{Q^2}{\mu^2}\right)^\ep\Biggl\{ 
\frac{1}{\ep^2} \frac{1}{2} P_{qg}^{(0)} P_{gq}^{(0)} 
+ \frac{1}{\ep} \left[ \frac{1}{2} P_{qq}^{(1),\rm PS} + P_{gq}^{(0)} \bar{c}^{(1)}_{g_1,g}\right] 
+ \bar{c}^{(2), \rm PS}_{g_1,q} + P_{qg}^{(0)} a_{g_1,g}^{(1)} \Biggr\},
\end{eqnarray}
\noindent
where  $\hat{a}_s = \hat{g}_s^2/(4\pi)^2$ denotes the unrenormalized strong coupling constant,
$S_\ep = \exp[\tfrac{\ep}{2}(\gamma_E - \ln(4\pi))]$ the spherical factor and $\gamma_E$ the 
Euler--Mascheroni constant,
$\bar{c}_i^{(k)}$ and $a_{g_1,g}^{(1)}$ are the expansion coefficients of the one-loop Wilson 
coefficient with
\begin{eqnarray}
\hat{\hat{C}}_{g_1,g}^{(1)} &=& \hat{a}_s S_\ep \left(\frac{Q^2}{\mu^2}\right)^{\ep/2} 
\left[ \frac{1}{\ep} P_{qg} + \bar{c}_{g_1,g}^{(1)} + \ep a_{g_1,g}^{(1)}\right],
\end{eqnarray}
given by the Feynman diagrams in Figure~\ref{fig:FIG1},
\noindent
where all quark lines are massless.
One obtains
\begin{eqnarray}
\bar{c}^{(1)}_{g_1,g} &=& 4 \textcolor{blue}{T_F N_F} \Bigl[- (2z-1)[\HA_1 + \HA_0] + 3 - 4z\Bigr]
\\
a_{g_1,g}^{(1)} &=& \textcolor{blue}{T_F N_F} \Bigl[
-12 + 16 z + 3 (1 - 2 z) \zeta_2 - (6 - 8 z) \left(\HA_0 + \HA_1\right) 
- (1 - 2 z) (\HA_0 + \HA_1)^2 \Bigr].
\end{eqnarray}
The contributing splitting functions \cite{Sasaki:1975hk,Ahmed:1976ee,Altarelli:1977zs,Mertig:1995ny,SP_PS1}
are
\begin{eqnarray}
P_{qg}(z) &=& 8 \textcolor{blue}{T_F N_F} \left[z^2 - (1-z)^2\right]
\\
P_{gq}(z) &=& 4 \textcolor{blue}{C_F} \frac{1-(1-z)^2}{z}
\\
P_{qq}^{(1),\rm PS}(z) &=& 16 \textcolor{blue}{C_F T_F N_F} \Bigl[ 1 - z - (1 - 3 z) \HA_0 - (1 + z) 
\HA_0^2\Bigr],
\end{eqnarray}
\noindent
where $C_F = (N_c^2-1)/{2 N_c}, T_F = 1/2$ and $N_c = 3$ for the gauge group $SU(3)_c$ of Quantum 
Chromodynamics. 
In the following we will use the convention
\begin{eqnarray}
\hat{f}(N_F) = f(N_F+1) - f(N_F).
\end{eqnarray}
The harmonic polylogarithms are given by \cite{Remiddi:1999ew}
\begin{equation}
\HA_{b,\vec{a}}(z) = \int_0^z dy f_b(y) \HA_{\vec{a}}(y),~~\HA_\emptyset = 1,~~b, a_i \in \{-1,0,1\},
\end{equation}
and the letters $f_c$ read
\begin{equation}\label{eq:HPL1}
f_0(z) = \frac{1}{z},~~~~
f_1(z) = \frac{1}{1-z},~~~~
f_{-1}(z) = \frac{1}{1+z}.
\end{equation}
We use the shorthand notation $\HA_{\vec{a}}(z) \equiv \HA_{\vec{a}}$ in case of the argument $z$. The harmonic 
polylogarithms are 
dual to the harmonic sums \cite{Vermaseren:1998uu,Blumlein:1998if} by the Mellin transformation (\ref{eq:MEL}).

\noindent
In the Larin--scheme we obtain
\begin{eqnarray}
\hat{\hat{C}}_{g_1}^{(2),\rm PS,L} &=& - \textcolor{blue}{C_F T_F N_F} \Biggl\{ 
-\frac{1}{\ep^2} \left[80 (1 - z) + 32 (1 + z) \HA_0\right]
+ \frac{1}{\ep} [184 (1 - z) - 32 (1 + z) \zeta_2 
\nonumber\\ &&
+ 40 (3 - z) \HA_0 
+ 24 (1 + z) \HA_0^2 + 80 (1 - z) \HA_1 + 32 (1 + z) \HA_{0,1}]
- \frac{1432}{3} (1-z)
\nonumber\\ &&
- \frac{4}{3} (233-43 z) \HA_0
+\frac{32 (1+z)^3}{3 z} \HA_{-1} \HA_0
-\frac{2}{3} \big(
        129-15 z+8 z^2\big) \HA_0^2
\nonumber\\ &&
-\frac{28}{3} (1+z) \HA_0^3
-(1-z)\big[
        184 
        +80 \HA_0
\big] \HA_1
-40 (1-z) \HA_1^2
-(1+z) \big[
        40 +
        32 \HA_0
\big] \HA_{0,1}
\nonumber\\ &&
-\frac{32 (1+z)^3}{3 z} \HA_{0,-1}
+16 (1+z) \left[\HA_{0,0,1}
- 2 \HA_{0,1,1} \right]
\nonumber\\ &&
+\left[
        \frac{4}{3} \big(
                129-45 z+8 z^2\big)
        +56 (1+z) \HA_0
\right] \zeta_2
+16 (1+z) \zeta_3,
\label{eq:C2PSunr}
\end{eqnarray}
performing the phase space integrations as has been outlined in \cite{Blumlein:2019qze,Zijlstra:1992qd}.

At $O(a_s^2)$ neither the renormalization of the heavy quark mass nor the coupling constant contributes
in case of the massive or massless pure singlet Wilson coefficient. The poles in $\ep$ in Eq.~(\ref{eq:C2PSunr}) 
are due to collinear singularities only, which have to be factorized. One may proceed as follows.
The unfactorized quarkonic Wilson coefficients for the structure function $g_1$, $\hat{\hat{C}}_{1,q}^{\rm NS,S}$
in Mellin--space are given by
\begin{eqnarray}
\hat{\hat{C}}_{1,q}^{\rm NS} &=& \Gamma_{qq}^{\rm NS} C_q^{\rm NS}
\\
\hat{\hat{C}}_{1,q}^{\rm S} &=& 
  \Gamma_{qq}^{\rm S} C_q^{\rm S}
+ \Gamma_{gq}^{\rm S} C_g^{\rm S}.
\end{eqnarray}
The pure singlet contribution is obtained by
\begin{eqnarray}
\hat{\hat{C}}_{1,q}^{\rm PS} &=& \hat{\hat{C}}_{1,q}^{\rm S} - \hat{\hat{C}}_{1,q}^{\rm NS}
\nonumber\\ &=& \Gamma_{qq}^{\rm S} C_q^{\rm S} - \Gamma_{qq}^{\rm NS} C_q^{\rm NS} + \Gamma_{gq}^{\rm S} 
C_g^{\rm S}
\\
&=& \left[\Gamma_{qq}^{\rm NS} + \Gamma_{qq}^{\rm S}\right]\left[C_q^{\rm NS} + C_q^{PS}\right] - 
\Gamma_{qq}^{\rm NS} C_q^{\rm NS} + \Gamma_{gq}^{\rm S} C_g^{\rm S}   
\end{eqnarray}
with
\begin{eqnarray} 
\Gamma_{gq}^{(0)} &=& \hat{a}_s S_\ep \left(\frac{\mu_F^2}{\mu^2}\right)^{\ep/2} 
\frac{1}{\ep} 
P_{gq}^{(0)}, \\ \Gamma_{qq}^{(1),\rm PS} &=& \hat{a}_s^2 S_\ep^2 \left(\frac{\mu_F^2}{\mu^2}\right)^{\ep} 
\left[
 \frac{1}{\ep^2} P_{qg}^{(0)} P_{gq}^{(0)} 
+\frac{1}{\ep} P_{qq}^{(1),\rm PS} \right].
\end{eqnarray}
and
\begin{eqnarray} 
\hat{\hat{C}}_{1,q}^{\rm PS} 
&=& a_s^2 \Biggl\{ 
  \frac{1}{\ep^2} \frac{1}{2} P_{qg}^{(0)}   P_{gq}^{(0)}
+ \frac{1}{\ep} \Biggl[\frac{1}{2} P_{qq}^{(1), \rm PS}  
+ P_{gq}^{(0)} C_g^{(1)} \Biggr] + C_q^{(2, \rm PS} \Biggr\} . 
\end{eqnarray}
The factorized massless pure singlet two--loop Wilson coefficient ${C}_{g_1}^{(2),\rm PS,L}$ is given by  
\begin{eqnarray}
C_{g_1}^{(2),\rm PS,L}\left(z,\frac{Q^2}{\mu^2}\right) &=& a_s^2 \Biggl\{
\frac{1}{8} P_{qg}^{(0)} P_{gq}^{(0)} L_M^2 + \frac{1}{2} 
\left[ P_{qq}^{(1),\rm PS} + P_{gq}^{(0)} \bar{c}_g^{(1)} \right] L_M + \bar{c}_q^{(2), \rm PS}  
\Biggr\},
\end{eqnarray}
where
\begin{eqnarray}
L_M = \ln\left(\frac{Q^2}{\mu}\right).
\end{eqnarray}
Here we set $\mu_F = \mu$, and work with a single scale only for the factorization and renormalization scale; 
$a_s = g_s^2/(4\pi)$ is the 
running coupling constant and the spherical factor $S_\ep$, as usually, is set to one at the end of the calculation. 
Note that the splitting function $P_{qq}^{(1),\rm PS}$ is correctly obtained, cf.~\cite{Mertig:1995ny,SP_PS1}, 
despite working in the Larin--scheme, cf. also \cite{Moch:2014sna}.

The massless Wilson coefficient in the {\sf M}--scheme is obtained by the following finite renormalization
\begin{eqnarray}
C_{g_1}^{(2),\rm PS, M} = C_{g_1}^{(2),\rm PS, L} - z_{\rm PS}^{(2)},
\end{eqnarray}
with \cite{Matiounine:1998re}
\begin{eqnarray}
\label{eq:ZPS2}
z_{\rm PS}^{(2)} = 
\textcolor{blue}{C_F T_F N_F} \Biggl[16 (1 - z) + 8 (3 - z) \HA_0 + 4 (2 + z) \HA^2_0 \Biggr],
\end{eqnarray}
cf.~\cite{Moch:2014sna,POL18}.
$C_{g_1}^{(2),\rm 
PS,M}$ in $z$--space is given by
\begin{eqnarray}
C_{g_1}^{(2),\rm PS,M}\left(z,\frac{Q^2}{\mu^2}\right) &=& a_s^2 \textcolor{blue}{C_F T_F N_F} \Biggl\{
\big[20 (1-z) +8 (1+z) \HA_0 \big] L_M^2
        - \big[
                (1-z) (88 + 40 \HA_1)
\nonumber\\ &&
                +16 (1+z) (\HA_0^2 + \HA_{0,1} - \zeta_2)
                +32 (2-z) \HA_0
           \big] L_M
     +   \frac{760}{3} (1-z)
\nonumber\\ &&
        +\frac{4}{3} (119-13 z) \HA_0
        -\frac{32 (1+z)^3}{3z}  \HA_{-1} \HA_0
        +\frac{2}{3} \big[
                75-15 z+8 z^2\big] \HA_0^2
\nonumber\\ &&
        +\frac{20}{3} (1+z) \HA_0^3
        +(1-z)\big(88 + 40  \HA_0\big)
        \HA_1
        +20 (1-z) \HA_1^2
        +\big[
                8 (3+z)
\nonumber\\ &&
                +16 (1+z) \HA_0
        \big] \HA_{0,1}
        +\frac{32 (1+z)^3}{3z} \HA_{0,-1}
        +16 (1+z) \HA_{0,1,1}
\nonumber\\ &&
        - 32\left[
                \frac{1}{3} \big(
                        9-3 z+z^2\big)
                + (1+z) \HA_0
        \right] \zeta_2
        -16 (1+z) \zeta_3
\Biggr\} .
\end{eqnarray}
\section{The Massive Wilson Coefficient}
\label{sec:4}

\vspace*{1mm}
\noindent
The kinematic domain for the massive Wilson coefficient is given by
\begin{eqnarray}
0 < z <  \frac{Q^2}{4m^2 + Q^2}.
\end{eqnarray}
The unrenormalized two--loop massive pure singlet Wilson coefficient reads in Mellin--$N$ space
\begin{eqnarray}
\hat{\hat{H}}_{g_1}^{(2),\rm PS,L} &=& \hat{a}_s^2 S_\ep^2 \left(\frac{Q^2}{\mu^2}\right)^\ep\Biggl\{
\frac{1}{\ep} P_{gq}^{(0)} h^{(1)}_{g_1,g}
+ h^{(2), \rm PS,L}_{g_1,q} + P_{qg}^{(0)} \bar{b}_{g_1,g}^{(1)} \Biggr\}.
\end{eqnarray}
The contributing Feynman diagrams are shown in Figure~\ref{fig:FIG2}, where now the outgoing quark lines
with momenta $k_1$ and $k_2$ are taken massive, see also Figure~\ref{fig:FIG1}.
Here $\bar{h}^{(1)}_{g_1,g}$ \cite{Watson:1981ce,Gluck:1990in,Vogelsang:1990ug} and $b_{g_1,g}^{(1)}$ are the 
expansion coefficients of the one--loop Wilson coefficient
\begin{eqnarray}
\hat{H}_{g_1,g}^{(1)} &=& \hat{a}_s S_\ep \left(\frac{Q^2}{\mu^2}\right)^{\ep/2} 
\left[ h_{g_1,g}^{(1)} + \ep \bar{b}_{g_1,g}^{(1)}\right] 
\end{eqnarray}
given by the diagrams in Figure~\ref{fig:FIG1} now with massive quark lines.
The expansion coefficients are given by 
\begin{eqnarray}
h^{(1)}_{g_1,g} &=& 4 T_F \left[(3-4z) \beta - (1-2z) \HA_0\left(\frac{1+\beta}{1-\beta}\right)\right]
\\
\bar{b}^{(1)}_{g_1,g} &=& 
T_F \Biggl\{
        -4 (3-4 z) \beta 
        +(1-2 z) \HA_0^2\left(
                \frac{1-\beta }{1+\beta }\right)
        -2 \left[(3-4 z) \beta 
                +(1-2 z) \HA_0\left(
                        \frac{1-\beta }{1+\beta }\right)
        \right]
\nonumber\\ && \times
\left[\HA_0 + \HA_1  - 2 \ln(\beta)\right]
        +4 (1-2 z) \HA_{0,1}\left(
                \frac{2 \beta }{1+\beta }\right)
\Biggr\}.
\end{eqnarray}
Here $\beta$ denotes the velocity of the produced heavy quarks,
\begin{eqnarray}
\beta = \sqrt{1 - \frac{4m^2}{Q^2} \frac{z}{1-z}}.
\end{eqnarray}
Since the two heavy quarks do not induce collinear divergences the mass factorization in the massive case reads
\begin{eqnarray}
        \hat{\hat{H}}_{g_1}^{(2),\rm PS,L} &=& H_{g_1}^{(2),\rm PS} + \Gamma_{gq} \otimes H_{g_1,g}^{(1),\rm 
PS} ~.
\end{eqnarray}
Here the Mellin convolution of two functions in $z$--space is given by
\begin{eqnarray}
\label{eq:MEL1}
A(z) \otimes B(z) = \int_0^1 dz_1 \int_0^1 dz_2 \delta(z - z_1 z_2) A(z_1) B(z_2).
\end{eqnarray}
We find
\begin{eqnarray}
        H_{g_1}^{(2),\rm PS,L} &=&  \hat{a}_s^2 S_\varepsilon^2 \biggl\{ \left( \frac{Q^2}{\mu^2} \right)^\ep 
\biggl[ \frac{1}{\ep} P_{gq}^{(0)} h_{g_1}^{(1)} + h_{g_1}^{(2),\rm PS} +  P_{gq}^{(0)} \bar{b}_{g_1}^{(1)} \biggr]
\nonumber \\ &&
- \left( \frac{\mu_F^2}{\mu^2} \right)^{\ep/2} \left( \frac{Q^2}{\mu^2} \right)^{\ep/2}  
\biggl[ \frac{1}{\ep} P_{gq}^{(0)} h_{g_1}^{(1)} + P_{gq}^{(0)} \bar{b}_{g_1}^{(1)} \biggr] \biggr\}.
\end{eqnarray} 
Identifying the renormalization and factorization scale, $\mu = \mu_F$, we finally obtain
\begin{eqnarray}
        H_{g_1}^{(2),\rm PS} &=&  {a}_s^2 \biggl[ \frac{1}{2} P_{gq}^{(0)} h_{g_1}^{(1)} L_M + h_{g_1}^{(2),\rm PS} 
\biggr] + O(\ep)~.
\end{eqnarray} 
Note that in the pure singlet case neither the heavy quark mass nor the coupling constant is  renormalized 
at two--loop order.

The massive pure singlet Wilson coefficient is obtained as a four--fold integral over two angular and two 
energy variables, cf.~\cite{Blumlein:2019qze} for details of the calculation. These integrals are 
systematically turned into iterative integrals. This process leads to a set of letters, through which 
these 
integrals are defined, see also \cite{Ablinger:2014bra}. We use the code \cite{RAAB1} in {\tt 
Mathematica}, 
which also uses the routine {\tt DSolveRational} of the package {\tt HolonomicFunctions} \cite{KOUTSCHAN}; 
see \cite{RAAB2, RAAB3} for the general theory underlying \cite{RAAB1}. We also refer to \cite{GUO} for the 
simpler case when no singularities are present at the endpoints of integration, which, however, does not 
apply here.

The result can be expressed in terms of iterative integrals of the following twelve partly square--root 
valued letters, 
\begin{eqnarray}
        f_{w_1}(t) &=& \frac{1}{1 - k t},
\\
        f_{w_2}(t) &=& \frac{1}{1 + k t},
\\
        f_{w_3}(t) &=& \frac{1}{\beta + t},
\\
        f_{w_4}(t) &=& \frac{1}{\beta - t},
\\
        f_{w_5}(t) &=& \frac{1}{k - z - ( 1 - z ) k t },
\\
        f_{w_6}(t) &=& \frac{1}{k + z - ( 1 - z ) k t },
\\
        f_{w_7}(t) &=& \frac{1}{k - z + ( 1 - z ) k t },
\\
        f_{w_8}(t) &=& \frac{1}{k + z + ( 1 - z ) k t },
\\
\label{eq:A1}
        f_{w_9}(t) &=& \frac{t}{k^2 \left(1 - t^2 \left(1 - z^2\right)\right)-z^2},
\\
        f_{w_{10}}(t) &=& \frac{1}{t \sqrt{1-t^2} \sqrt{1-k^2 t^2}},
\\
        f_{w_{11}}(t) &=& \frac{t}{\sqrt{1-t^2} \sqrt{1-k^2 t^2}},
\\
\label{eq:AA1}
        f_{w_{12}}(t) &=& \frac{t}{\sqrt{1-t^2} \sqrt{1-k^2 t^2} \left(k^2 \left(1 - t^2 \left(1 -
z^2\right)\right)-z^2\right)},
\end{eqnarray}
given before in Ref.~\cite{Blumlein:2019qze}, and the letters spanning the harmonic polylogarithms. These 
iterative integrals have maximally 
weight {\sf w = 3} and belong to the  Kummer--elliptic integrals, \cite{Blumlein:2019qze}, in general.
The variable $k$ is defined by
\begin{eqnarray}
k = \frac{\sqrt{z}}{\sqrt{1-(1-z)\beta^2}}.
\end{eqnarray}
One obtains for the following analytic result of the massive polarized two--loop Wilson coefficient 
\begin{eqnarray}
H_{g_1}^{(2),\rm PS,L} &=& \textcolor{blue}{C_F T_F} \Biggl\{
-\frac{16(1-z) P_1}{3 k^2}
\biggl\{
        \HA_{w_{5},0}
      - \HA_{w_{6},0}
      + \HA_{w_{7},0}
      - \HA_{w_{8},0}
      -\bigl[
              \HA_{w_{5}}
            - \HA_{w_{6}}
            + \HA_{w_{7}}
            - \HA_{w_{8}}
      \bigr] 
\nonumber \\ &&
\times \HA_0
\biggr\}
-\frac{8 P_2}{3 k^2} \HA_{w_{2},-1}
-\frac{8 P_3}{3 k^2} \HA_1 \HA_{w_{1}}
+ \frac{4(1-z) P_4}{3 k^2 z}
\biggl\{
        \HA_{w_{6},1}
      - \HA_{w_{8},-1}
      - \HA_{w_{6}} \HA_1
\nonumber \\ &&
      + \HA_{w_{8}} \HA_{-1}
\biggr\}
- \frac{4 (1-z) P_5}{3 k^2 z}
\biggl\{
        \HA_{w_{6},-1}
      - \HA_{w_{8},1}
      + \HA_{w_{8}} \HA_1
      - \HA_{w_{6}} \HA_{-1}
\biggr\}
\nonumber \\ &&
+\frac{4 (1-z) P_6}{3 k^2 z}
\biggl\{
        \HA_{w_{5},-1}
      - \HA_{w_{7},1}
      + \HA_{w_{7}} \HA_1
      - \HA_{w_{5}} \HA_{-1}
\biggr\} 
-\frac{4 (1-z) P_7}{3 k^2 z}
\biggl\{
        \HA_{w_{5},1}
\nonumber \\ &&
      - \HA_{w_{7},-1}
      - \HA_{w_{5}} \HA_1
      + \HA_{w_{7}} \HA_{-1}
\biggr\}
+ \frac{16 P_8}{3(1- k \beta)} \HA_{w_{1}}
- \frac{16 P_9}{3(1+k \beta)} \HA_{w_{2}}
\nonumber \\ &&
+ \frac{8 (1-z) P_{10}}{3 (k (2-z)-z) (1-k \beta)} \HA_{w_{5}}
+ \frac{8 (1-z) P_{11}}{3 (k (2-z)+z) (1+k \beta)} \HA_{w_{6}}
\nonumber \\ &&
- \frac{8 (1-z) P_{12}}{3 (k (2-z)-z) (1+k \beta)} \HA_{w_{7}}
- \frac{8 (1-z) P_{13}}{3 (k (2-z)+z) (1-k \beta)} \HA_{w_{8}}
\nonumber \\ &&
+ \frac{32 P_{14}}{3 k^2 \big(k^2 (2-z)^2-z^2\big)} \HA_1
- \frac{32 P_{15}}{3 k^2 \big(k^2 (2-z)^2-z^2\big)} \HA_{-1}
\nonumber \\ &&
+\frac{1216}{3} (1-z) \beta
+ 8 (1-z) (1-2z)
\bigl[
            \HA_1 + \HA_{-1} - 2 \beta 
\bigr] \big( \ln(z) + \ln(1-z) \big)
\nonumber \\ &&
+ 16 (1 + 2z)
\biggl\{
        2 \bigl( \HA_{w_{1},w_{4}} + \HA_{w_{2},w_{4}} + \HA_{w_{3},w_{1}} + \HA_{w_{3},w_{2}} \bigr)
\nonumber \\ &&
      + k \bigl( \HA_{w_{1}}^2 - \HA_{w_{2}}^2 \bigr)
      + \bigl[
            - 2 \ln \big(k^2-z\big)
            + 6 \ln (k)
            - \ln \big(1-k^2\big)
            + \ln (k^2-z^2)
            - 2 \HA_{w_{3}}
      \bigr] 
\nonumber \\ && \times
      \bigl( \HA_{w_{1}} + \HA_{w_{2}} \bigr)
      + k ( 1 - z ) \bigl[ \HA_{w_{5},w_{1}} + \HA_{w_{6},w_{2}} - \HA_{w_{7},w_{2}} - \HA_{w_{8},w_{1}} \bigr]
\nonumber \\ &&
      - k ( 1 - z ) \bigl[ \HA_{w_{5}} - \HA_{w_{8}} \bigr] \HA_{w_{1}}
      - k ( 1 - z ) \bigl[ \HA_{w_{6}} - \HA_{w_{7}} \bigr] \HA_{w_{2}}
\biggr\}
\nonumber \\ &&
+16 (1-z) (7-2 z) \beta  \ln (k^2-z^2)
+ 8 \left(7-\left(2-\frac{1}{k^2}\right) z\right)
\biggl\{
      2 \HA_1 \HA_0
      - 6 \ln (k) \HA_1
\nonumber \\ &&
      + \ln \big(1-k^2\big) \HA_1
      + 2 \ln \big(k^2-z\big) \HA_1
      + 2 \HA_{w_{3}} \HA_1
      - 6 \ln (k) \HA_{-1}
      + \ln \big(1-k^2\big) \HA_{-1}
\nonumber \\ &&
      + 2 \ln \big(k^2-z\big) \HA_{-1}
      + 2 \HA_{-1} \HA_{w_{3}}
      - 2 \HA_{0,1}
      - 2 \HA_{1,w_{4}}
      - 2  \HA_{w_{3},1}
      - 2 \HA_{w_{3},-1}
      + 2  \HA_{-1,0}
\nonumber \\ &&
      - 2 \HA_{-1,w_{4}}
\biggr\}
+\frac{32}{3} \big(-3 k^2+\frac{z^2}{k^2}-3 \big(1+z^2\big)\big) \bigl[ \HA_{w_{1}} + \HA_{w_{2}} \bigr] \HA_0
+8 \ln (k^2-z^2) 
\nonumber \\ && \times
\left(-8+\frac{\big(-1+5 k^2\big) z}{k^2}-2 z^2\right) \bigl[ \HA_1 + \HA_{-1} \bigr]
+\frac{8}{3} \big(6 k^2-\frac{3 z}{k}
-\frac{2 z^2}{k^2}+6 \big(2+2 z
\nonumber \\ &&
+z^2\big)-3 k \big(8-5 z+2 z^2\big)\big) \bigl[ \HA_{w_{1}} \HA_{-1} - \HA_{w_{2}} \HA_1 \bigr]
+\frac{8}{3} \big(6 k^2+\frac{3 z}{k}-\frac{2 z^2}{k^2}+6 \big(2+2 z
\nonumber \\ &&
+z^2\big)+3 k \big(8-5 z+2 z^2\big)\big) \HA_{w_{2}} \HA_{-1}
+ \frac{4}{3} \left(87+\frac{4}{z}-\frac{9 \big(-1+2 k^2\big) z}{k^2}\right) \bigl[ \HA_1^2 - \HA_{-1}^2 \bigr]
\nonumber \\ &&
-\frac{16}{3} k (1-z) \big(3+2 z+3 z^2\big)
\biggl\{
       \frac{1}{\sqrt{z}}
      \bigl[
              \HA_{w_{10},w_{5}}
            - \HA_{w_{10},w_{6}}
            + \HA_{w_{10},w_{7}}
            - \HA_{w_{10},w_{8}}
      \bigr]
\nonumber \\ &&
      - \frac{k}{\sqrt{z}}
      \biggl(
              \HA_{w_{5},w_{11}}
            + \HA_{w_{6},w_{11}}
            + \HA_{w_{7},w_{11}}
            + \HA_{w_{8},w_{11}}
      -\bigl[
              \HA_{w_{5}}
            + \HA_{w_{6}}
            + \HA_{w_{7}}
            + \HA_{w_{8}}
      \bigr] \HA_{w_{11}}
      \biggr)
\nonumber \\ &&
      - 2 k ( 1 - k^2 ) \sqrt{z}
      \biggl(
              \HA_{w_{5},w_{12}}
            + \HA_{w_{6},w_{12}}
            + \HA_{w_{7},w_{12}}
            + \HA_{w_{8},w_{12}}
      - \bigl[
              \HA_{w_{5}}
            + \HA_{w_{6}}
            + \HA_{w_{7}}
\nonumber \\ &&
            + \HA_{w_{8}}
      \bigr] \HA_{w_{12}}
      \biggr)
      - \frac{2}{\sqrt{z}(1-z)} \bigl[ \HA_{w_{10},w_{1}} + \HA_{w_{10},w_{2}} \bigr]
      - \frac{2 (1 - k^2) \sqrt{z}}{1-z} \bigl[ \HA_{w_{12},1} 
\nonumber \\ &&
+ \HA_{w_{12},-1} \bigr]
\biggr\}
-\frac{384 \big(k^2-z\big)}{k^2 \beta } \HA_{w_{3}}
- \frac{8}{3} \left(39-\frac{4}{z}+\frac{9 \big(1-2 k^2\big) z}{k^2}\right) \HA_{-1} \HA_1
\nonumber \\ &&
+ 32 \biggl( k^2 - (2-z) z-\frac{z^2}{3 k^2}\biggr) \HA_{w_{1},0}
+\frac{8}{3} \bigl(6 k^2+\frac{3 z}{k}-\frac{2 z^2}{k^2}-3 k z (1-2 z)
\nonumber \\ &&
-6 \big(1+4 z-z^2\big)\bigr) \HA_{w_{1},1}
+\frac{8}{3} \big(6-6 k^2+24 z+\frac{3 z}{k}-6 z^2+\frac{2 z^2}{k^2}
\nonumber \\ &&
-3 k z (1-2 z)\big) \HA_{w_{1},-1}
+32 \biggl( k^2 - (2 - z) z -\frac{z^2}{3 k^2}\biggr) \HA_{w_{2},0}
+\frac{8}{3} \biggl(6 k^2-\frac{3 z}{k}
\nonumber \\ &&
+3 k (1-2 z) z-\frac{2 z^2}{k^2}
-6 \big(1+4 z-z^2\big)\biggr) \HA_{w_{2},1}
+ \biggl( 208-\frac{64}{3 z}+\frac{48 \big(1-2 k^2\big) z}{k^2} \biggr) 
\nonumber \\ &&
\times \HA_{-1,1}
- \frac{64 k^2 (1-z^2) \big(1+3 z^2\big)}{3 z}
\biggl\{
      \HA_{w_{9},1} 
      + \HA_{w_{9},-1}
      - k (1-z) \bigl[ \HA_{w_{9},w_{5}} + \HA_{w_{9},w_{6}} 
\nonumber \\ &&
      + \HA_{w_{9},w_{7}} + \HA_{w_{9},w_{8}} \bigr]
\biggr\}
+ 8 (1+z)
\biggl\{
      - 4 \HA_{0,1,1}
      - 4 \HA_{0,-1,1}
      + 20 \HA_{1,1,1}
      + 4 \HA_{1,1,w_{4}}
\nonumber \\ &&
      + 4 \HA_{1,-1,w_{4}}
      - 4 \HA_{w_{3},1,1}
      + 4 \HA_{w_{3},1,-1}
      - 4 \HA_{w_{3},-1,1}
      + 4 \HA_{w_{3},-1,-1}
      + 4 \HA_{-1,1,0}
\nonumber \\ &&
      + 16 \HA_{-1,1,1}
      - 4 \HA_{-1,1,w_{4}}
      + 4 \HA_{-1,-1,0}
      + 16 \HA_{-1,-1,1}
      - 4 \HA_{-1,-1,w_{4}}
      + 20 \HA_{-1,-1,-1}
\nonumber \\ &&
+ \biggl(        
         \ln \big(1-k^2\big)
        - \ln (k^2-z^2)
        +2 \ln \big(k^2-z\big)
        -6  \ln (k)
\biggr) \bigl[ \HA_{-1}^2 - \HA_1^2 - 2 \HA_1 \HA_{-1} 
\nonumber \\ &&
      + 4 \HA_{-1,1} \bigr]
+\bigl[
         \bigl(    
                10 \HA_{-1}          
                -4 \HA_{w_{3}}
        \bigr) \HA_1
        -4 \HA_0 \HA_1
        +2 \HA_1^2
        -2 \HA_{1,1}
        -4 \HA_{w_{3},1}
        -4 \HA_{w_{3},-1}
\nonumber \\ &&
        -12 \HA_{-1,1}
        -10 \HA_{-1,-1}
\bigr] \HA_{-1}
+\bigr[
         4 \HA_{1,1}
        +8 \HA_{-1,1}
        +4 \HA_{-1,-1}
        - 4 \HA_1^2
\bigl] \HA_{w_{3}}
+\bigl[
         4 \HA_{0,1}
\nonumber \\ &&
        +4 \HA_{0,-1}
        -10 \HA_{1,1}
        +4 \HA_{w_{3},1}
        +4 \HA_{w_{3},-1}
        -4 \HA_{-1,1}
        -10 \HA_{-1,-1}
\bigr] \HA_1
\nonumber \\ &&
+\bigl[       
        -4 \HA_1^2
        +4 \HA_{1,1}
        +4 \HA_{-1,1}
\bigr] \HA_0
\biggr\}
+ 32 k (1+z)
\biggl\{
         \HA_{w_{1},1,0}
      +  \HA_{w_{1},1,1}
      -  \HA_{w_{1},1,w_{4}}
\nonumber \\ &&
      -  \HA_{w_{1},1,-1}
      +  \HA_{w_{1},-1,0}
      +  \HA_{w_{1},-1,1}
      -  \HA_{w_{1},-1,w_{4}}
      -  \HA_{w_{1},-1,-1}
      -  \HA_{w_{2},1,0}
      -  \HA_{w_{2},1,1}
\nonumber \\ &&
      +  \HA_{w_{2},1,w_{4}}
      +  \HA_{w_{2},1,-1}
      -  \HA_{w_{2},-1,0}   
      -  \HA_{w_{2},-1,1}
      + \HA_{w_{2},-1,w_{4}}
      + \HA_{w_{2},-1,-1}
      + \HA_{w_{3},1,w_{1}}
\nonumber \\ &&
      - \HA_{w_{3},1,w_{2}}
      + \HA_{w_{3},-1,w_{1}}
      - \HA_{w_{3},-1,w_{2}}
      + \frac{1}{2} 
      \bigl[
             \ln (k^2-z^2)
            - 2 \ln \big(k^2-z\big)
            + 6 \ln (k)
\nonumber \\ &&
            - \ln \big(1-k^2\big)
      \bigr] \bigl[ \HA_{w_{2},-1} - \HA_{w_{1},-1} + \HA_{w_{2},1} - \HA_{w_{1},1} \bigr]
+ \frac{1}{2} \biggl(
        - \HA_{w_{1},1}
        - \HA_{w_{1},-1}
        + \HA_{w_{2},1}
\nonumber \\ &&
        + \HA_{w_{2},-1}
        + \bigl[ \HA_{w_{1}} - \HA_{w_{2}} \bigr] \HA_1
\biggr) \HA_{-1}
+\bigl[
          \HA_{w_{1},1}
        + \HA_{w_{1},-1}
        - \HA_{w_{2},1}
        - \HA_{w_{2},-1}
\bigr] \HA_{w_{3}}
\nonumber \\ &&
- \frac{1}{2} \biggl(
          \HA_{1,1}
        - \HA_1^2
        + 2 \HA_{w_{3},1}
        + 2 \HA_{w_{3},-1}
        + 2 \HA_{-1,1}
        + \HA_{-1,-1}
        + k \HA_{w_{1},1}
        + k \HA_{w_{1},-1}
\nonumber \\ &&
        - k \HA_{w_{2},1}
        - k \HA_{w_{2},-1}
\biggr) \bigl[ \HA_{w_{1}} - \HA_{w_{2}} \bigr]
+\frac{1}{2}\bigl[
          \HA_{w_{1},1}
        + \HA_{w_{1},-1}
        - \HA_{w_{2},1}
        - \HA_{w_{2},-1}
\bigr] \HA_1
\biggr\}
\nonumber \\ &&
- 16 k ( 1 - z^2 )
\biggl\{
- \HA_{1,w_{4},w_{5}}
- \HA_{1,w_{4},w_{6}}
- \HA_{1,w_{4},w_{7}}
- \HA_{1,w_{4},w_{8}}
+ \HA_{w_{5},1,1}
- \HA_{w_{5},1,-1}
\nonumber \\ &&
+ \HA_{w_{5},w_{3},1}
- \HA_{w_{5},w_{3},-1}
+ \HA_{w_{6},1,1}
- \HA_{w_{6},1,-1}
+ \HA_{w_{6},w_{3},1}
- \HA_{w_{6},w_{3},-1}
+ \HA_{w_{7},w_{3},1}
\nonumber \\ &&
- \HA_{w_{7},w_{3},-1}
- \HA_{w_{7},-1,1}
+ \HA_{w_{7},-1,-1}
+ \HA_{w_{8},w_{3},1}
- \HA_{w_{8},w_{3},-1}
- \HA_{w_{8},-1,1}
+ \HA_{w_{8},-1,-1}
\nonumber \\ &&
+ \HA_{-1,w_{4},w_{5}}
+ \HA_{-1,w_{4},w_{6}}
+ \HA_{-1,w_{4},w_{7}}
+ \HA_{-1,w_{4},w_{8}}
+ k \bigl[
 \HA_{w_{1},w_{4},w_{5}}
+\HA_{w_{1},w_{4},w_{6}}
\nonumber \\ &&
+\HA_{w_{1},w_{4},w_{7}}
+\HA_{w_{1},w_{4},w_{8}}
-\HA_{w_{2},w_{4},w_{5}}
-\HA_{w_{2},w_{4},w_{6}}
-\HA_{w_{2},w_{4},w_{7}}
-\HA_{w_{2},w_{4},w_{8}}
-\HA_{w_{5},1,w_{1}}
\nonumber \\ &&
+\HA_{w_{5},1,w_{2}}
-\HA_{w_{5},w_{3},w_{1}}
+\HA_{w_{5},w_{3},w_{2}}
-\HA_{w_{6},1,w_{1}}
+\HA_{w_{6},1,w_{2}}
-\HA_{w_{6},w_{3},w_{1}}
+\HA_{w_{6},w_{3},w_{2}}
\nonumber \\ &&
-\HA_{w_{7},w_{3},w_{1}}
+\HA_{w_{7},w_{3},w_{2}}
+\HA_{w_{7},-1,w_{1}}
-\HA_{w_{7},-1,w_{2}}
-\HA_{w_{8},w_{3},w_{1}}
+\HA_{w_{8},w_{3},w_{2}}
\nonumber \\ &&
+\HA_{w_{8},-1,w_{1}}
-\HA_{w_{8},-1,w_{2}}
\bigr]
+\biggl(
        - \HA_{1,1}
        - \HA_{w_{3},1}
        + \HA_{w_{3},-1}
        - \HA_{-1,1}
        + \HA_1^2
        - k \HA_{w_{1},1}
\nonumber \\ &&
        + k \HA_{w_{2},1}
        + k \HA_{w_{3},w_{1}}
        - k \HA_{w_{3},w_{2}}
\biggr) \bigl[ \HA_{w_{5}} + \HA_{w_{6}} \bigr]
+\biggl(
        - \HA_{w_{3},1}
        + \HA_{w_{3},-1}
        + \HA_{-1,1}
\nonumber \\ &&
        + \HA_{-1,-1}
        - \HA_1 \HA_{-1}
        + k \HA_{w_{1},-1}
        - k \HA_{w_{2},-1}
        + k \HA_{w_{3},w_{1}}
        - k \HA_{w_{3},w_{2}}
\biggr) \bigl[ \HA_{w_{7}} + \HA_{w_{8}} \bigr]
\nonumber \\ &&
+\biggl(
        \bigl(                
                  \HA_{w_{5}}
                + \HA_{w_{6}}
                + \HA_{w_{7}}
                + \HA_{w_{8}}
        \bigr) \HA_{w_{3}}
        -  \HA_{w_{5},1}
        -  \HA_{w_{5},w_{3}}
        -  \HA_{w_{6},1}
        -  \HA_{w_{6},w_{3}}
        -  \HA_{w_{7},w_{3}}
\nonumber \\ &&
        +  \HA_{w_{7},-1}
        -  \HA_{w_{8},w_{3}}
        +  \HA_{w_{8},-1}
\biggr) 
\nonumber \\ && \times
\bigl[ \HA_1 -  \HA_{-1} - k \bigl( \HA_{w_{1}} - \HA_{w_{2}} \bigr) \bigr]
\biggr\}
+ 576 (1-z) \beta  \ln (k)
- 96 (1-z) \beta  \ln \big(1-k^2\big)
\nonumber \\ &&
- 192 (1-z) \beta \HA_0
- 192 (1-z) \beta  \ln \big(k^2-z\big)
\Biggr\}
\nonumber \\ &&
+ \frac{1}{2} P_{gq}^{(0)} \otimes h_{g_1}^{(1)} L_M
- P_{gq}^{(0)} \otimes \bar{b}_{g_1}^{(1)}.
\label{eq:g1full}
\end{eqnarray}
The remaining convolutions appearing in Eq.~(\ref{eq:g1full}) are given in Appendix~\ref{sec:A1}.
Here and in the Appendix the argument of the iterative integrals $\HA_{\vec{a}}$ is $\beta$.

The polynomials $P_i$ in Eq.~(\ref{eq:g1full}) read 
\begin{eqnarray} 
P_1&=&3 k^4+3 k^2 \left(z^2+1\right)-z^2, \\ 
P_2&=&6 k^4+3 k^3 z (2 z-1)+6 k^2 \left(z^2-4 z-1\right)+3 k z-2 z^2, \\ 
P_3&=&6 k^4+3 k^3 \left(2 z^2-5 z+8\right)+6 k^2 \left(z^2+2 z+2\right)+3 k z-2 
z^2, \\ 
P_4&=&6 k^4 z+k^3 \left(-16 z^3+33 z^2-24 z+8\right)+6 k^2 z \left(z^2+2 z+2\right)-3 k z^2-2 z^3, \\ 
P_5&=&6 k^4 z+k^3 \left(-4 z^3+3 z^2+24 z+8\right)+6 k^2 z \left(z^2+2 z+2\right)+3 k z^2-2 z^3, \\ 
P_6&=&6 
k^4 z+k^3 \left(4 z^3-3 z^2-24 z-8\right)+6 k^2 z \left(z^2+2 z+2\right)-3 k z^2-2 z^3, \\ P_7&=&6 k^4 z+k^3 
\left(16 z^3-33 z^2+24 z-8\right)+6 k^2 z \left(z^2+2 z+2\right)+3 k z^2-2 z^3, \\ P_8&=&3 \beta k^3 (z-3)+3 
k^2 (z-4)+4 \beta k \left(3 z^2-13 z+3\right)-12 z^2+46 z+9, \\ P_9&=&3 \beta k^3 (z-3)-3 k^2 (z-4)+4 \beta k 
\left(3 z^2-13 z+3\right)+12 z^2-46 z-9, \\ P_{10}&=&3 \beta k^4 (1-z)^2+k^3 \bigl(-35 \beta +(3-16 \beta ) 
z^2+3 (17 \beta -9) z+39\bigr)+k^2 \bigl(-52 \beta +12 \beta z^3 \nonumber \\ && +(22-81 \beta ) z^2+(121 
\beta -72) z+35\bigr)+k \bigl(12 (\beta -1) z^3+(75-38 \beta ) z^2 \nonumber \\ && +(26 \beta -88) 
z+10\bigr)+z \bigl(-12 z^2+32 z-5\bigr), \\ P_{11}&=&3 \beta k^4 (1-z)^2+k^3 \bigl(35 \beta +(16 \beta -3) 
z^2+(27-51 \beta ) z-39\bigr)+k^2 \bigl(-52 \beta +12 \beta z^3 \nonumber \\ && +(22-81 \beta ) z^2+(121 
\beta -72) z+35\bigr)+k \bigl(-12 (\beta -1) z^3+(38 \beta -75) z^2 \nonumber \\ && +(88-26 \beta ) 
z-10\bigr)+z \bigl(-12 z^2+32 z-5\bigr), \\ P_{12}&=&3 \beta k^4 (1-z)^2-k^3 \bigl(35 \beta +(16 \beta +3) 
z^2-3 (17 \beta +9) z+39\bigr)+k^2 \bigl(-52 \beta +12 \beta z^3 \nonumber \\ && -(81 \beta +22) z^2+(121 
\beta +72) z-35\bigr)+k \bigl(12 (\beta +1) z^3-(38 \beta +75) z^2 \nonumber \\ && +(26 \beta +88) 
z-10\bigr)+z \bigl(12 z^2-32 z+5\bigr), \\ P_{13}&=&3 \beta k^4 (1-z)^2+k^3 \bigl(35 \beta +(16 \beta +3) 
z^2-3 (17 \beta +9) z+39\bigr)+k^2 \bigl(-52 \beta +12 \beta z^3 \nonumber \\ && -(81 \beta +22) z^2+(121 
\beta +72) z-35\bigr)+k \bigl(-12 (\beta +1) z^3+(38 \beta +75) z^2 \nonumber \\ && -2 (13 \beta +44) 
z+10\bigr)+z \bigl(12 z^2-32 z+5\bigr), \\ P_{14}&=&k^4 \bigl(-3 (36 \beta +1)+(27 \beta -10) z^3+(37-135 
\beta ) z^2+(216 \beta -34) z\bigr) \nonumber \\ && +k^2 z \bigl((3-27 \beta ) z^2+(27 \beta +28) 
z-28\bigr)+7 z^3, \\ P_{15}&=&k^4 \bigl(-108 \beta +(27 \beta +10) z^3-(135 \beta +37) z^2+(216 \beta +34) 
z+3\bigr) \nonumber \\ && +k^2 z \bigl(-3 (9 \beta +1) z^2+(27 \beta -28) z+28\bigr)-7 z^3.
\end{eqnarray} 
We finally perform the transformation to the {\sf M}--scheme for the massive two--loop pure singlet Wilson 
coefficient. It is given by 
\begin{eqnarray} 
H_{g_1}^{2,\rm PS,M} = H_{g_1}^{2,\rm PS,L} - \frac{1}{N_F} z_{\rm PS}^{(2)}. 
\end{eqnarray} 
\section{The Asymptotic and Threshold Expansions}
\label{sec:5}

\vspace*{1mm}
\noindent
The complete expressions calculated in Section~\ref{sec:4} allow now to perform the asymptotic expansion
for $Q^2 \gg m^2$ and the threshold expansion for $\beta \ll 1$.

In the asymptotic limit $Q^2 \gg m^2$ the first expansion coefficients of the polarized massive pure singlet 
Wilson coefficient read setting $\mu^2 = Q^2$
\begin{eqnarray}
        H_{g_1}^{(2),\rm PS,M} &=& \textcolor{blue}{C_F T_F} \Biggl\{
-\big(
        20 (1-z)
        + 8 (1+z) \HA_0
\big) \ln^2 \left( \frac{\textcolor{blue}{m^2}}{\textcolor{blue}{Q^2}} \right)
-\big(
        8 (1-z)
        -8 (1-3 z) \HA_0
\nonumber \\ &&
        -8 (1+z) \HA_0^2
\big)  \ln \left( \frac{\textcolor{blue}{m^2}}{\textcolor{blue}{Q^2}} \right)
+\frac{592}{3} (1-z)
+\biggl(
         \frac{256}{3} (2-z)
        -\frac{32 (1+z)^3}{3 z} \HA_{-1}
\biggr) \HA_0
\nonumber \\ &&
+\frac{8}{3} \big(21+2 z^2\big) \HA_0^2
+\frac{16}{3} (1+z) \HA_0^3
+\biggl(
         88 (1-z)
        +80 (1-z) \HA_0
\biggr) \HA_1
+20 (1-z) \HA_1^2
\nonumber \\ &&
-\biggl(
        16 (1-3 z)
        -32 (1+z) \HA_0
\biggr) \HA_{0,1}
+\frac{32 (1+z)^3}{3 z} \HA_{0,-1}
-32 (1+z) \HA_{0,0,1}
\nonumber \\ &&
+16 (1+z) \HA_{0,1,1}
-\big(
         \frac{32}{3} \big(9-3 z+z^2\big)
        +32 (1+z) \HA_0
\big) \zeta_2
+16 (1+z) \zeta_3
\nonumber \\ &&
- \Biggl[16 (1 - z) + 8 (3 - z) \HA_0 + 4 (2 + z) \HA^2_0 \Biggr]
\nonumber \\ &&
+ \frac{\textcolor{blue}{m^2}}{\textcolor{blue}{Q^2}}
\Biggl[
\bigl(
        16 (1-z) (1-3 z)
        -32 z \HA_0
\bigr) \ln \left( \frac{\textcolor{blue}{m^2}}{\textcolor{blue}{Q^2}} \right)
+8 \bigl(18-12 z-7 z^2\bigr)
\nonumber \\ &&
+16 \bigl(6+z+6 z^2\bigr) \HA_0
+16 z \HA_0^2
+16 \bigl(3-7 z+3 z^2\bigr) \HA_1
\Biggr]
\nonumber \\ &&
+ \left(\frac{\textcolor{blue}{m^2}}{\textcolor{blue}{Q^2}}\right)^2
\Biggl[
-4 (1-z) (3+4 z) \ln^2 \left( \frac{\textcolor{blue}{m^2}}{\textcolor{blue}{Q^2}} \right)
+\biggl(
         \frac{4 P_{18}}{1-z}
        -16 (1-z) (5+4 z) \HA_0
\nonumber \\ &&
        -8 (1-z) (5+4 z) \HA_1
\biggr) \ln \left( \frac{\textcolor{blue}{m^2}}{\textcolor{blue}{Q^2}} \right)
+\frac{2 P_{19}}{3 (1-z)^2}
+\biggl(
         \frac{16 P_{16}}{1-z}
        -64 (1-z^2) \HA_{-1}
\biggr) \HA_0
\nonumber \\ &&
+\biggl(
         \frac{4 P_{17}}{1-z}
        -32 (1-z) \HA_0
\biggr) \HA_1
-4 (1-z) (7+4 z) \HA_1^2
-16 (1-z) (3+4 z) \HA_{0,1}
\nonumber \\ &&
+64 (1-z^2) \HA_{0,-1}
+16 (1-z) \zeta_2
\Biggr]
+ O \left( \kappa^3 \ln^2(\kappa) \right)
\Biggr\}, 
\label{eq:ASY}
\end{eqnarray}
with the polynomials
\begin{eqnarray}
P_{16} &=& 3 z^4+z^3-11 z^2+13 z-7,
\\
P_{17} &=& 6 z^4+2 z^3-63 z^2+84 z-32,
\\
P_{18} &=& 6 z^4+2 z^3-57 z^2+76 z-28,
\\
P_{19} &=& 15 z^5-27 z^4+393 z^3-1079 z^2+1069 z-339.
\end{eqnarray}
In this expansion the Kummer--elliptic integrals turn into harmonic polylogarithms.
The leading term, which is free of power corrections of $O((m^2/Q^2)^k), k \in \mathbb{N}, k \geq 1$, 
can be predicted using the representation of the massive Wilson coefficient by massive operator matrix 
elements (OMEs), cf.~\cite{Buza:1995ie,Buza:1996xr,Bierenbaum:2007qe,POL18}, and massless Wilson 
coefficients,
\begin{eqnarray}
H_{g_1}^{(2),\rm PS}\left(z,\frac{Q^2}{m^2}\right) &=&  A_{Qq}^{(2),\rm PS}(N_F+1) + \hat{C}_{g_1}^{(2),\rm 
PS}(N_F+1).
\end{eqnarray}  
Here the massless Wilson coefficient $\tilde{C}_{g_1}^{(2),\rm PS}(N_F+1)$ is the one given in 
Section~\ref{sec:3} normalized by $N_F+1$. The massive two--loop operator matrix element  
$A_{Qq}^{(2),\rm PS}$ in Mellin space reads
\begin{eqnarray}
A_{Qq}^{(2),\rm PS} &=& -\frac{1}{8} \hat{P}_{qg}^{(0)} P_{gq}^{(0)} \ln^2\left(\frac{m^2}{\mu^2}\right) 
- \frac{1}{2} \hat{P}_{qq}^{(1), \rm PS} \ln\left(\frac{m^2}{\mu^2}\right) 
+ \frac{1}{8} \hat{P}_{qg}^{(0)} P_{gq}^{(0)} \zeta_2 + a_{Qq}^{(2),\rm PS},
\label{eq:asymp}
\end{eqnarray}  
cf.~\cite{Buza:1995ie,Buza:1996xr,Bierenbaum:2007qe,POL18}; for its renormalization see 
Ref.~\cite{Bierenbaum:2009mv}. The constant part of the unrenormalized polarized OME $a_{Qq}^{(2),\rm PS}$ 
is given by \cite{Buza:1996xr,POL18}
\begin{eqnarray}
a_{Qq}^{(2),\rm PS}(z) &=& \textcolor{blue}{C_F T_F} \Biggl\{
- 72 (1-z)
-12 (1+5 z) \HA_0
-2 (1-3 z) \HA_0^2
-\frac{4}{3} (1+z) \HA_0^3
+40 (1-z) \HA_0 \HA_1
\nonumber\\ &&
-\big(
        40 (1-z)
       -16 (1+z) \HA_0
\big) \HA_{0,1}
-32 (1+z) \HA_{0,0,1}
-\big(
        20 (1-z) -
        8 (1+z) \HA_0
\big) \zeta_2
\nonumber\\ &&
+32 (1+z) \zeta_3
\Biggr\}  
\end{eqnarray}  
in $z$-space. The calculation of $A_{Qq}^{(2),\rm PS}$ is performed in the Larin--scheme. One has to apply
the tensor decomposition method, however, to obtain the correct result. These aspects are discussed in 
Ref.~\cite{POL18} in detail.

The threshold expansion of the Wilson coefficients for $\beta \ll 1$ is given 
by
\begin{eqnarray}
H_{g_1}^{(1)}\left(z,\frac{Q^2}{m^2}\right) &=& 4 \textcolor{blue}{T_F} \beta 
\biggl\{ 1
-\frac{2}{3} (1-2 z) \beta ^2
-\frac{2}{5} (1-2 z) \beta ^4
-\frac{2}{7} (1-2 z) \beta ^6
\\ \nonumber &&
-\frac{2}{9} (1-2 z) \beta ^8
+ O(\beta^{10})
\biggr\},
\\
H_{g_1}^{(2),\rm PS,L}\left(z,\frac{Q^2}{m^2}\right) &=& \textcolor{blue}{C_F T_F} (1-z) \beta^3
\Biggl\{
-\frac{256}{9}
+\frac{16}{3} \bigl[ \ln (1-z) - \ln (z) + 4 \ln (2 \beta ) \bigr]
\\ \nonumber &&
+\beta ^2 \biggl(
        -\frac{32}{75} (41+20 z)
        +\frac{16}{5} \bigl[ \ln (1-z) - \ln (z) + 4 \ln (2 \beta ) \bigr]
\biggr)
\\ \nonumber &&
-\beta ^4 \biggl(
        \frac{16 \big(2723+20504 z-12352 z^2\big)}{11025}
        -\frac{16}{105} \big(7+16 z-8 z^2\big) \bigl[ \ln (1-z) 
\\ \nonumber &&
- \ln (z) + 4 \ln (2 \beta ) \bigr]
\big)
+\beta ^6 \biggl(
         \frac{16 \big(47203-909904 z+950864 z^2-345728 z^3\big)}{297675}
\\ \nonumber &&
        +\frac{16}{945} \big(1+272 z-232 z^2+64 z^3\big) \bigl[ \ln (1-z) - \ln (z) + 4 \ln (2 \beta \biggr) 
\biggr]
\biggr)
\\ \nonumber &&
+ O ( \beta^8 )
\Biggr\}.
\end{eqnarray}

\section{Numerical Results}
\label{sec:6}

\begin{figure}[h!]
\centering
\includegraphics[width=0.7 \linewidth]{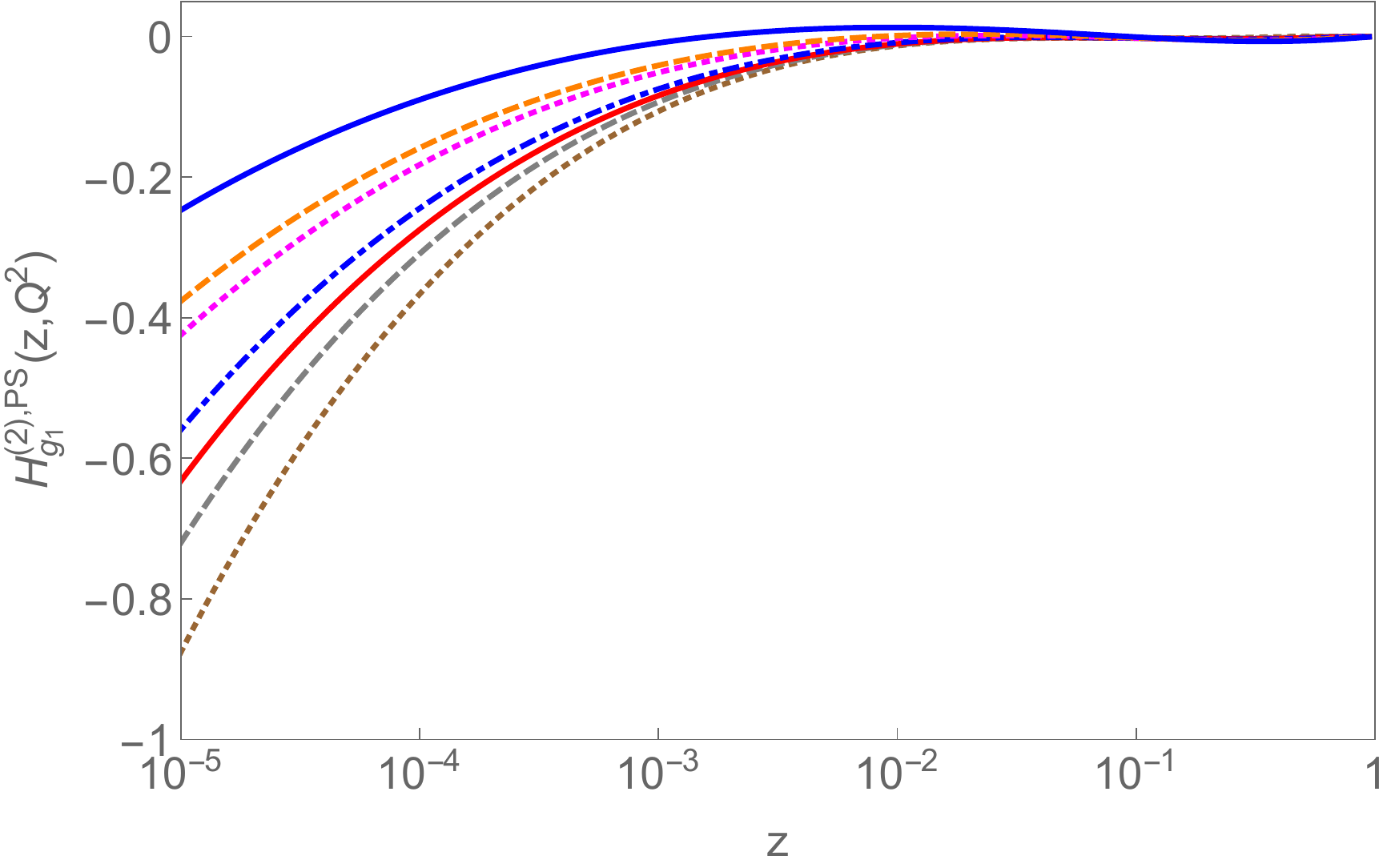}
\caption{\small \sf The Wilson coefficient $H_{g_{1}}^{(2),\rm PS}$ 
as a function of $z$ for 
different values of $Q^2$ and the scale choice $\mu^2 = \mu_F^2 = Q^2$. 
Upper full line (Blue):      $Q^2=10^4~\GeV^2$; 
upper dashed line (Orange):  $Q^2=10^3~\GeV^2$; 
upper dotted line (Magenta): $Q^2=500~\GeV^2$; 
dash-dotted line (Blue):     $Q^2=100~\GeV^2$; 
lower full line (Red):       $Q^2=50~\GeV^2$; 
lower dashed line (Gray):    $Q^2=25~\GeV^2$; 
lower dotted line (Brown):   $Q^2=10~\GeV^2$.}
\label{fig:F1}
\end{figure}

\vspace*{1mm}
\noindent
Let us now illustrate the analytic results numerically. In Figure~\ref{fig:F1} the two--loop heavy flavor
Wilson coefficient $H_{g_1}^{(2),\rm PS,M}$ is shown  as a function of $z$ for different values of $Q^2 
\in [10, 10^4]~\GeV^2$, setting the charm quark mass to $m_c = 1.59~\GeV$, cf.~\cite{Ablinger:2014vwa}.
\begin{figure}[H]\centering
\includegraphics[width=0.7\linewidth]{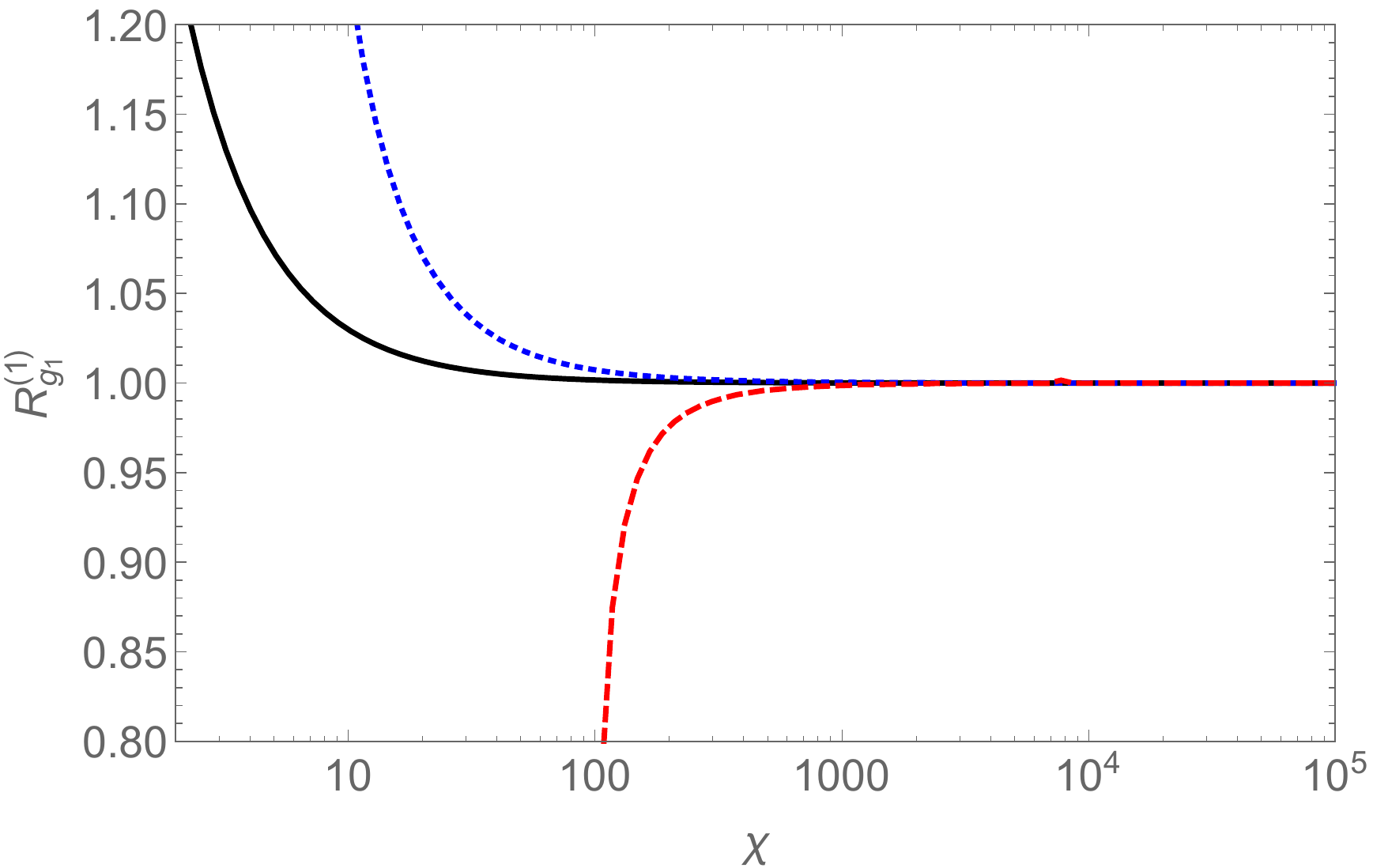}
\caption{\sf \small The ratio $R_{g_1}^{(1)}$, Eq.~(\ref{eq:RAT1}),  as a 
function 
of $\chi=Q^2/m^2$. Solid line: $z=10^{-4}$; dotted line: $z=10^{-2}$; dashed line: $z=1/2$.}
\label{fig:RAT1}
\end{figure}
\noindent
For large values of $Q^2$ these results approach the asymptotic result for $H_{g_1,q}^{2,\rm PS}$.
In the small $x$ region this Wilson coefficient is negative.

Next we study the ratios
\begin{eqnarray}
\label{eq:RAT1}
        R_{g_1}^{(1)} &=& \frac{H_{g_1,q}^{2,\rm PS}}{\tilde{H}_{g_1,q}^{2,\rm PS}} (\mu =\mu_F =m)~,
\end{eqnarray}
comparing the full (\ref{eq:g1full}) and the asymptotic results, $\tilde{H}$,
(\ref{eq:ASY}) for the leading term in Figure~\ref{fig:RAT1}.

\begin{figure}[H]
\includegraphics[width=0.49 \linewidth]{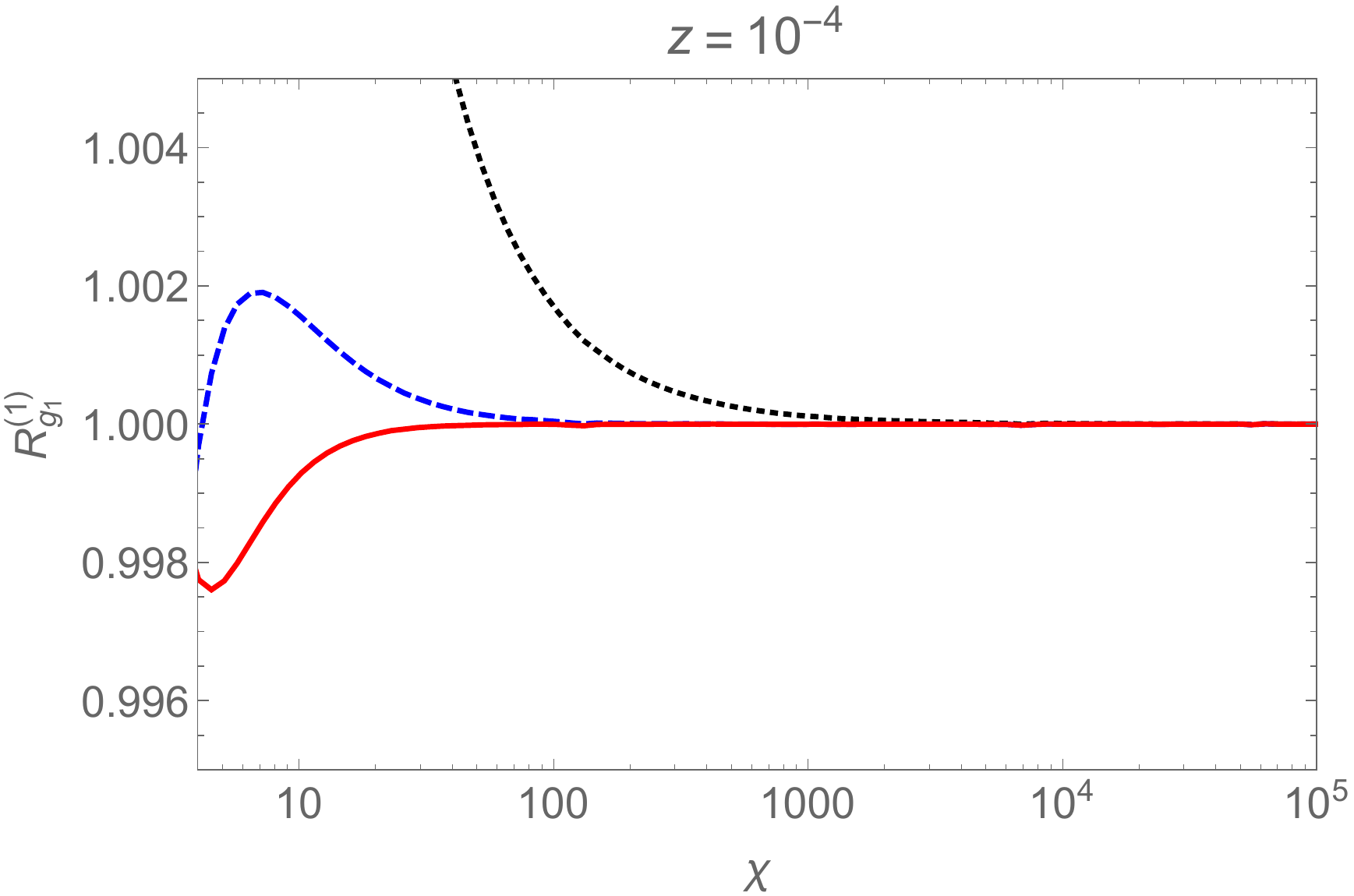}
\includegraphics[width=0.49 \linewidth]{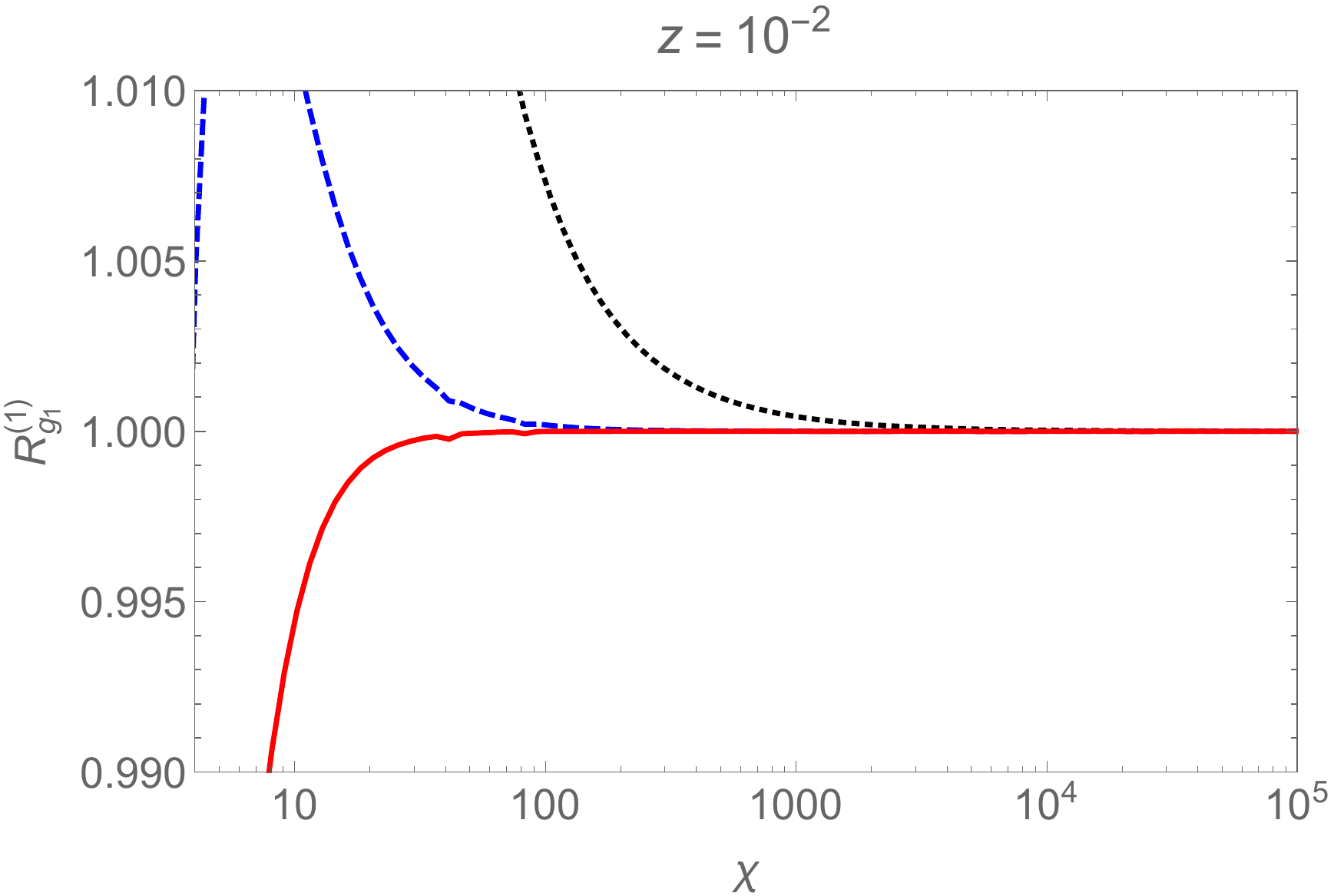}
\includegraphics[width=0.49 \linewidth]{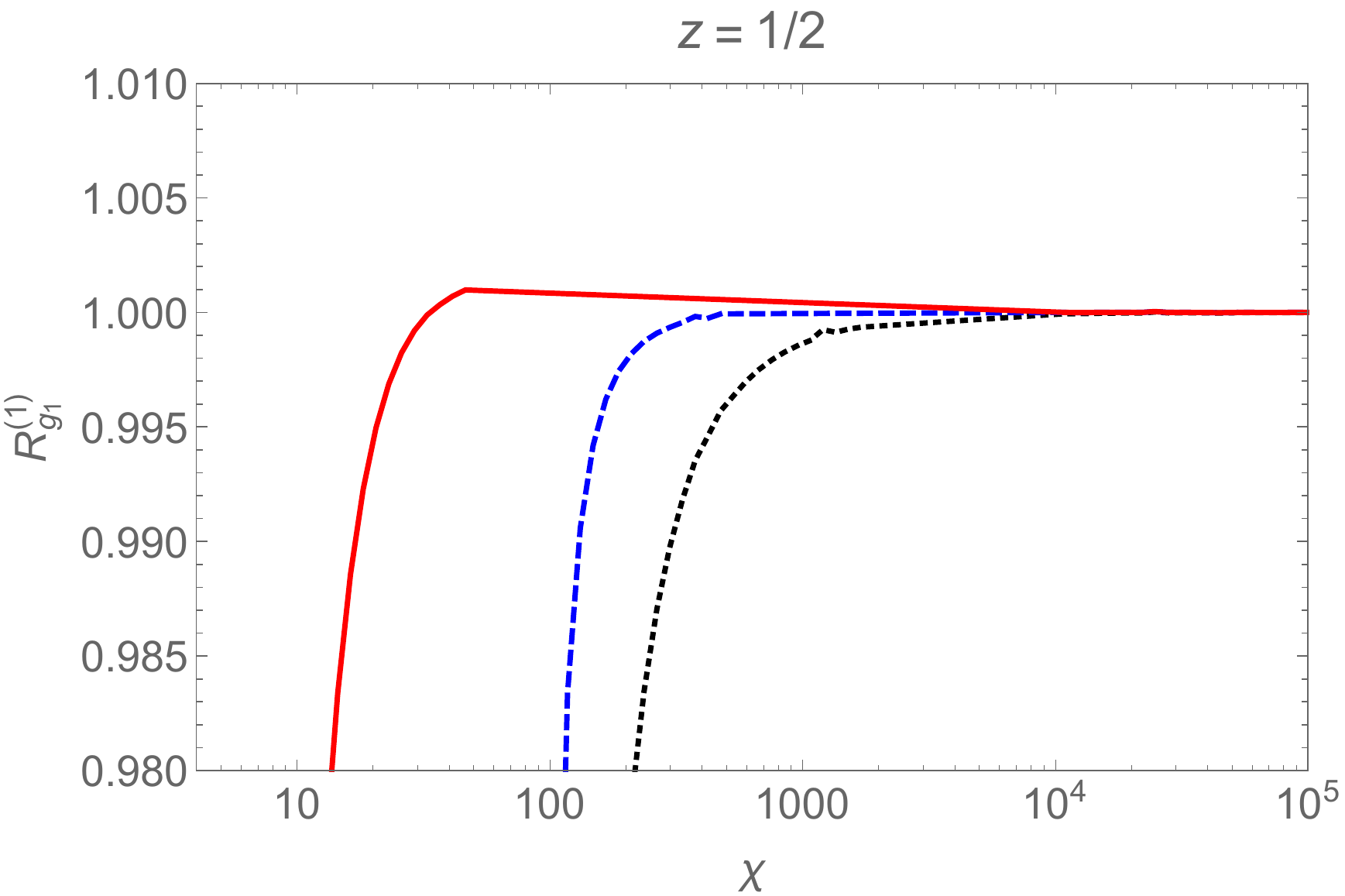}
\caption{\small \sf The ratio $R_{g_1}^{(1)}$, Eq.~(\ref{eq:RAT1}),                    
as a function of $\chi=Q^2/m^2$ for different values of $z$ gradually improved with $\kappa$ suppressed terms.
Dotted lines: asymptotic result; dashed lines: $O(m^2/Q^2)$ improved; solid lines : $O((m^2/Q^2)^2)$ improved.}
\label{fig:RLandR2b}
\end{figure}

For $H_{g_1,q}^{2,\rm PS}$ the asymptotic expansion agrees with the full calculation 
up to $Q^2/m^2 \equiv \chi = 10$ to about $2\%$ for $z=10^{-4}$, $\chi = 40$ for $z = 10^{-2}$ and $\chi = 
200$ for $z = 1/2$. However, the Wilson coefficients are very small already for the last value.
Similar to the ratio of the full and asymptotic Wilson coefficient we define the ratio 
\begin{eqnarray}
\label{eq:RAT2}
        R_{g_{1}} &=& \frac{ g_{1,q}^{(2), \rm PS} }{ \tilde{g}_{1,q}^{(2), \rm PS} },
\end{eqnarray}
where $\tilde{g}_{1,q}^{(2), \rm PS}$ is the structure function obtained by using the expansion of the
respective Wilson coefficient up to the desired level. The corresponding results are depicted in 
Figure~\ref{fig:RLandR2b}. We use the parameterization of the parton distributions Ref.~\cite{Blumlein:2010rn} at 
NLO with the corresponding values of $\alpha_s(Q^2)$ at NNLO \cite{Alekhin:2017kpj} to compare to previous
non--singlet results in \cite{Behring:2015zaa}. Demanding an agreement within $\pm 2\%$ for $g_1^{\rm PS}$ in the 
range $z \in [10^{-4}, 10^{-2}, 1/2]$ leads to values $Q_0^2/m^2 \in [5, 5, 13]$ of the $O((m^2/Q^2)^2)$ 
improved result, $Q_0^2/m^2 \in [10, 12, 30]$ of the $O(m^2/Q^2)$ improved 
result, and $Q_0^2/m^2 \in [12, 100, 170]$ for the asymptotic result. 

\begin{figure}[H]
\centering
\includegraphics[width=0.49 \linewidth]{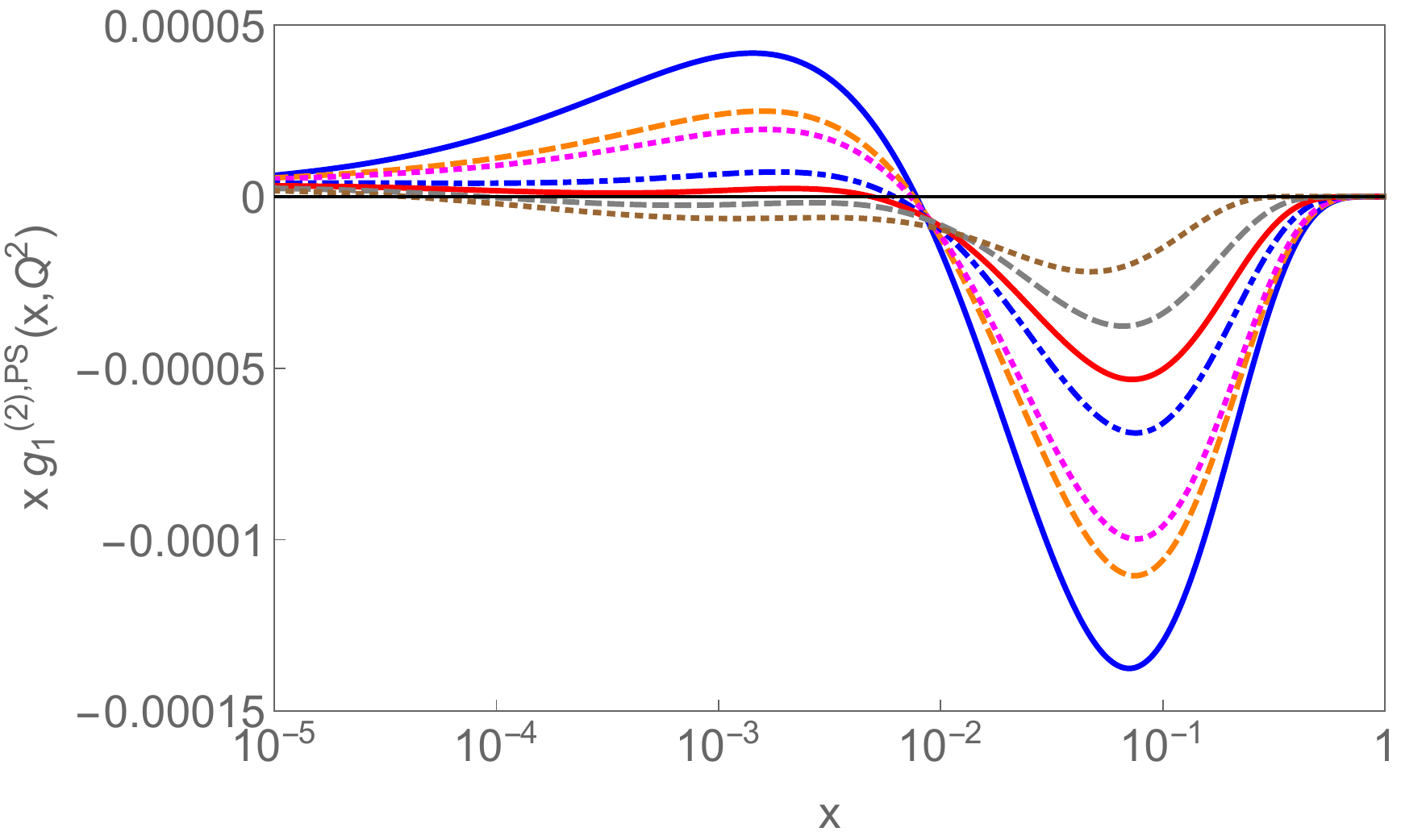}
\includegraphics[width=0.49 \linewidth]{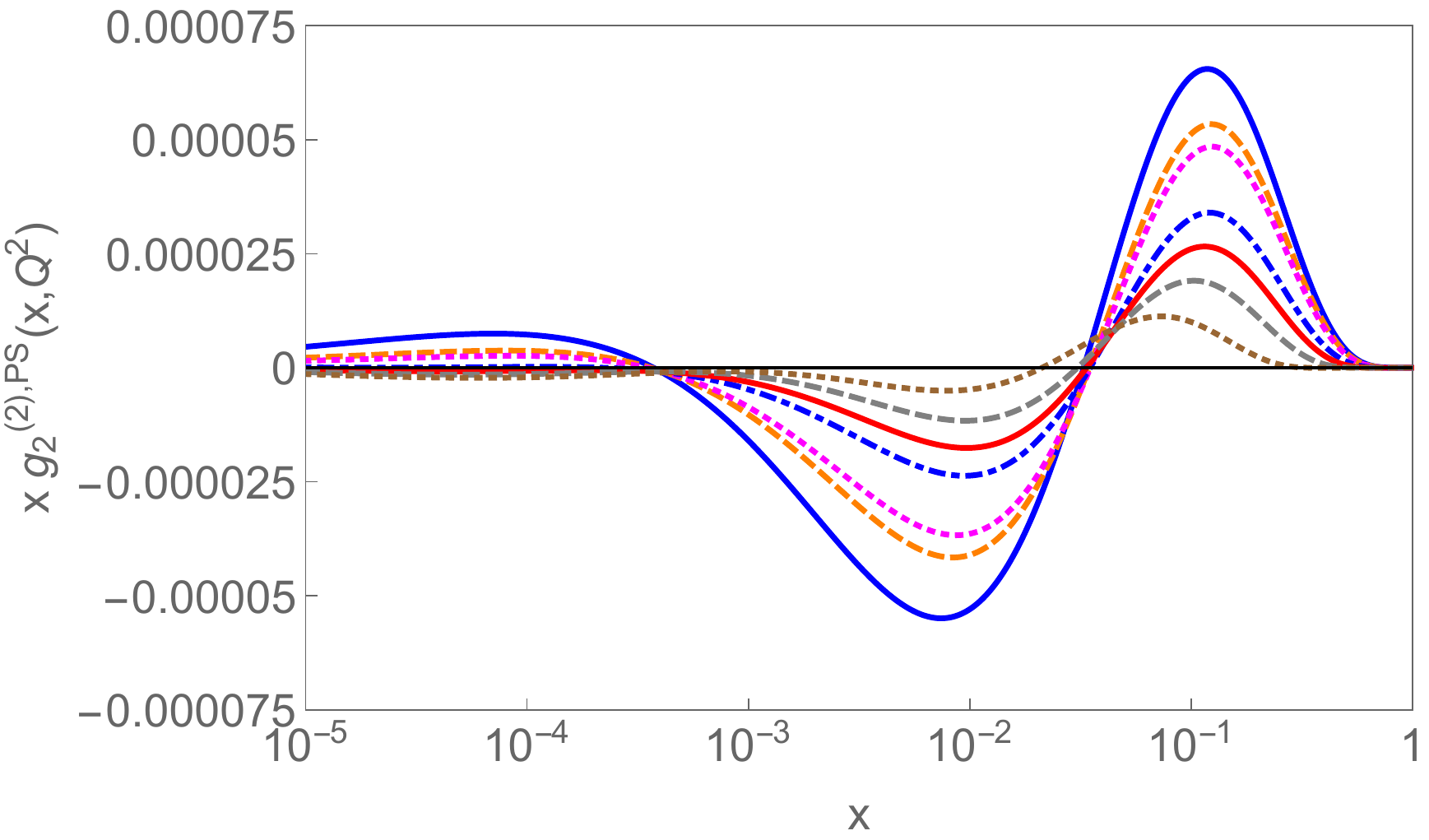}
\caption{\small \sf The pure singlet contributions 
$xg_{1}^{(2),\rm PS}$ 
and
$xg_{2}^{(2),\rm PS}$ 
for different values of $Q^2$ and the scale choice $\mu^2 = \mu_F^2 = Q^2$. 
Full line (Blue):        $Q^2=10^4~\GeV^2$; 
dashed line (Orange):    $Q^2=10^3~\GeV^2$; 
dotted line (Magenta):   $Q^2=500~\GeV^2$; 
dash-dotted line (Blue): $Q^2=100~\GeV^2$; 
full line (Red):         $Q^2=50~\GeV^2$; 
dashed line (Gray):      $Q^2=25~\GeV^2$; 
dotted line (Brown):     $Q^2=10~\GeV^2$, 
using the parameterization of the parton distribution \cite{Blumlein:2010rn}.}
\label{fig:FIG4}
\end{figure}

In Figures~\ref{fig:FIG4} we show the complete results for the two--loop pure singlet contributions to $xg_1$ 
and $xg_2$ as a function of $x$ for a series of $Q^2$--values. Both functions show an oscillatory 
behaviour,
which is enlarged for $xg_2$ due to the Wandzura--Wilczek relation.
\begin{figure}[H]
\centering
\includegraphics[width=0.49 \linewidth]{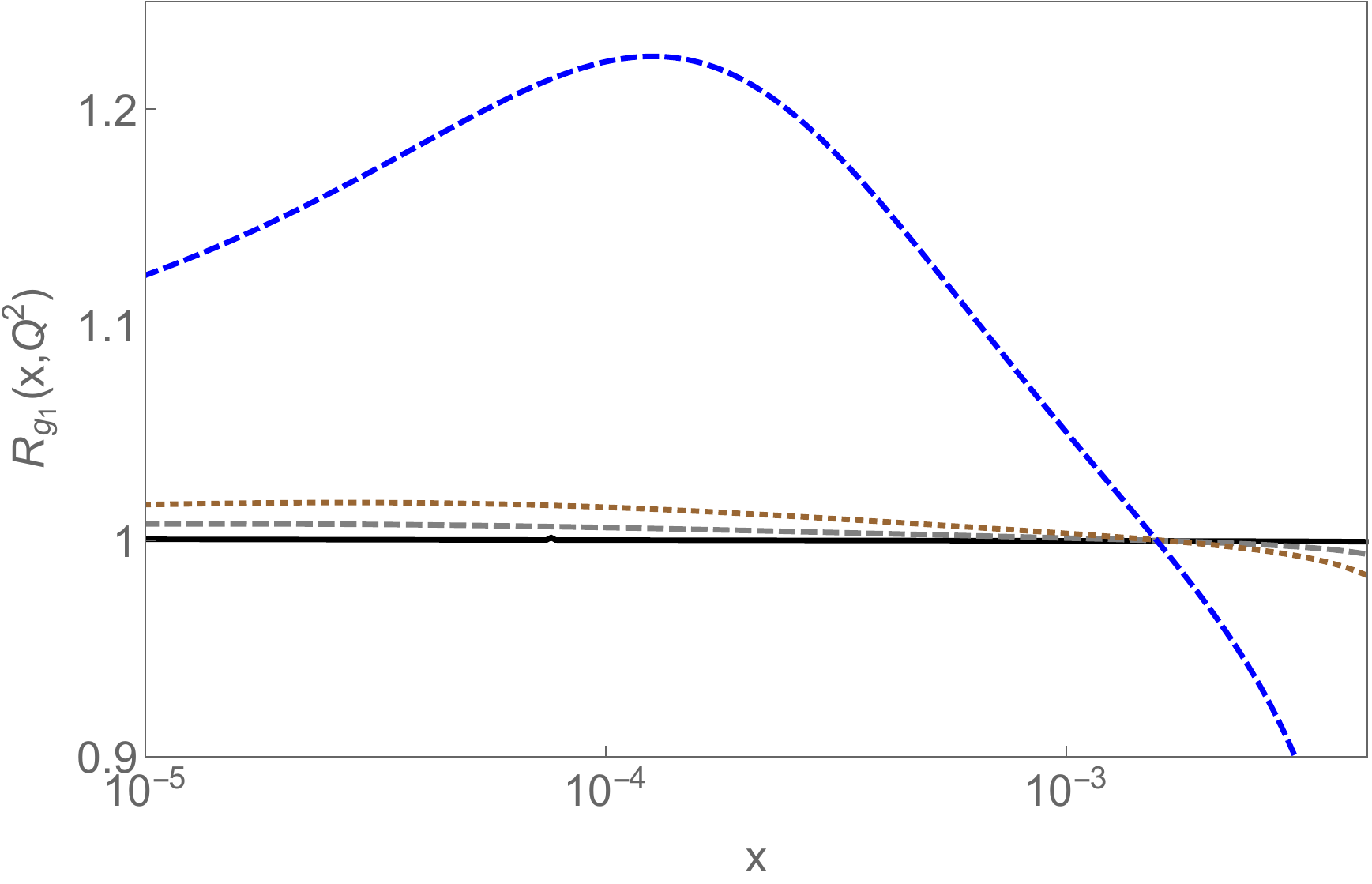}
\includegraphics[width=0.49 \linewidth]{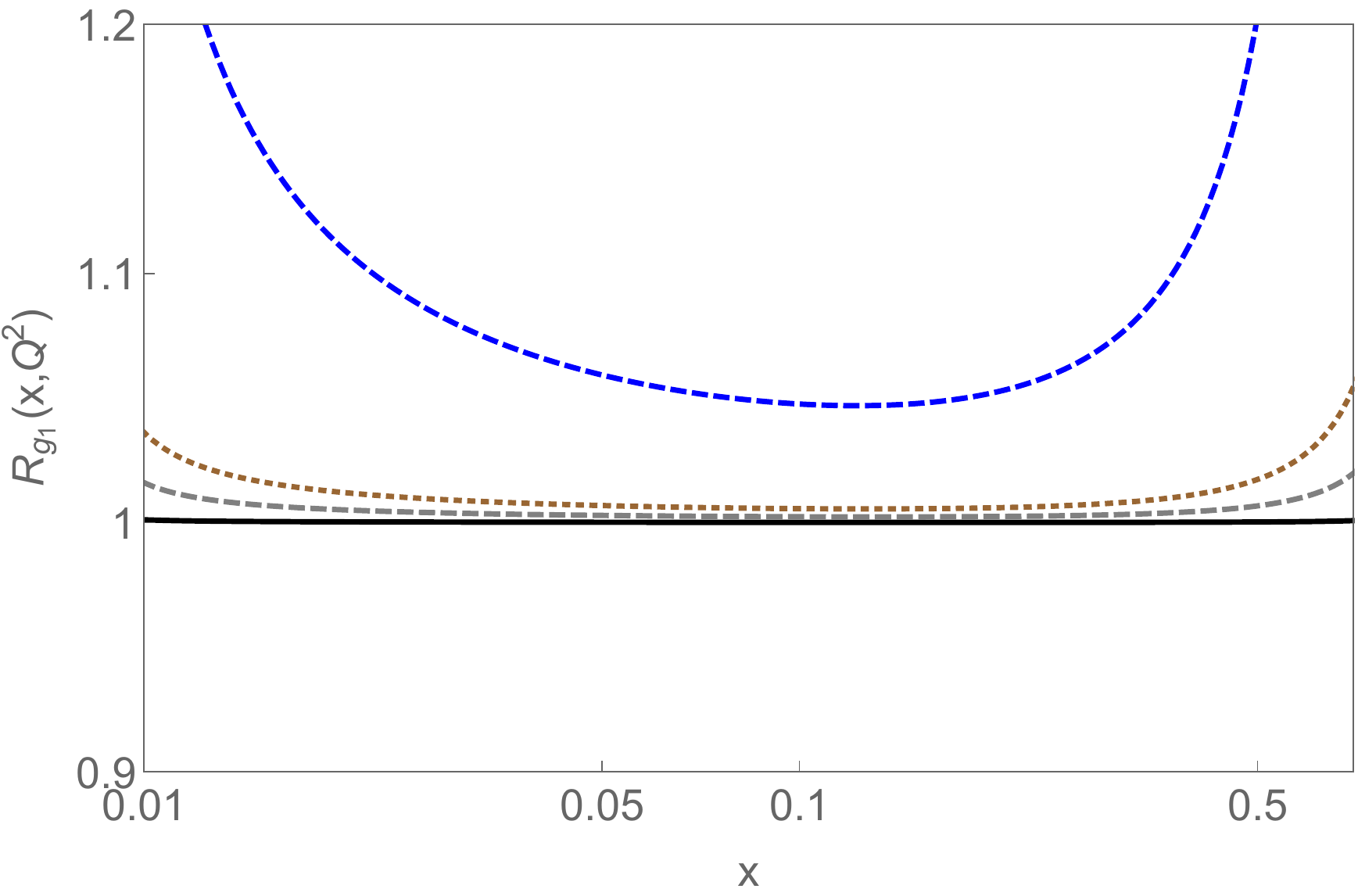}
\caption{\small \sf The ratio of the structure function  $g_1^{(2),\rm PS}$ 
in the full calculation over the asymptotic approximation for different values of $Q^2$ and the scale choice $\mu^2 = \mu_F^2 = 
Q^2$. 
Full line (Black):        $Q^2=10^4~\GeV^2$; 
dashed line (Gray):       $Q^2=10^3~\GeV^2$; 
dotted line (Brown):      $Q^2=500~\GeV^2$;
lower dashed line (Blue): $Q^2=100~\GeV^2$,
using the parameterization of the parton distribution \cite{Blumlein:2010rn}}.
\label{fig:FIG5}
\end{figure}
In Figure~\ref{fig:FIG5} we illustrate the ratios Eq.~(\ref{eq:RAT2}) as a function of $x$ for different values 
of $Q^2$ for $g_1^{\rm PS}$ comparing the asymptotic result to the full result. For a better visibility and to avoid 
to depict zero transitions in the denominator we separate the small $x$ and large $x$ part into two plots.
The corrections behave widely flat in $x$ for larger values of $Q^2$ and develop some profile for $Q^2 < 
100~\GeV^2$.

\begin{figure}[H]
\centering
\includegraphics[width=0.49 \linewidth]{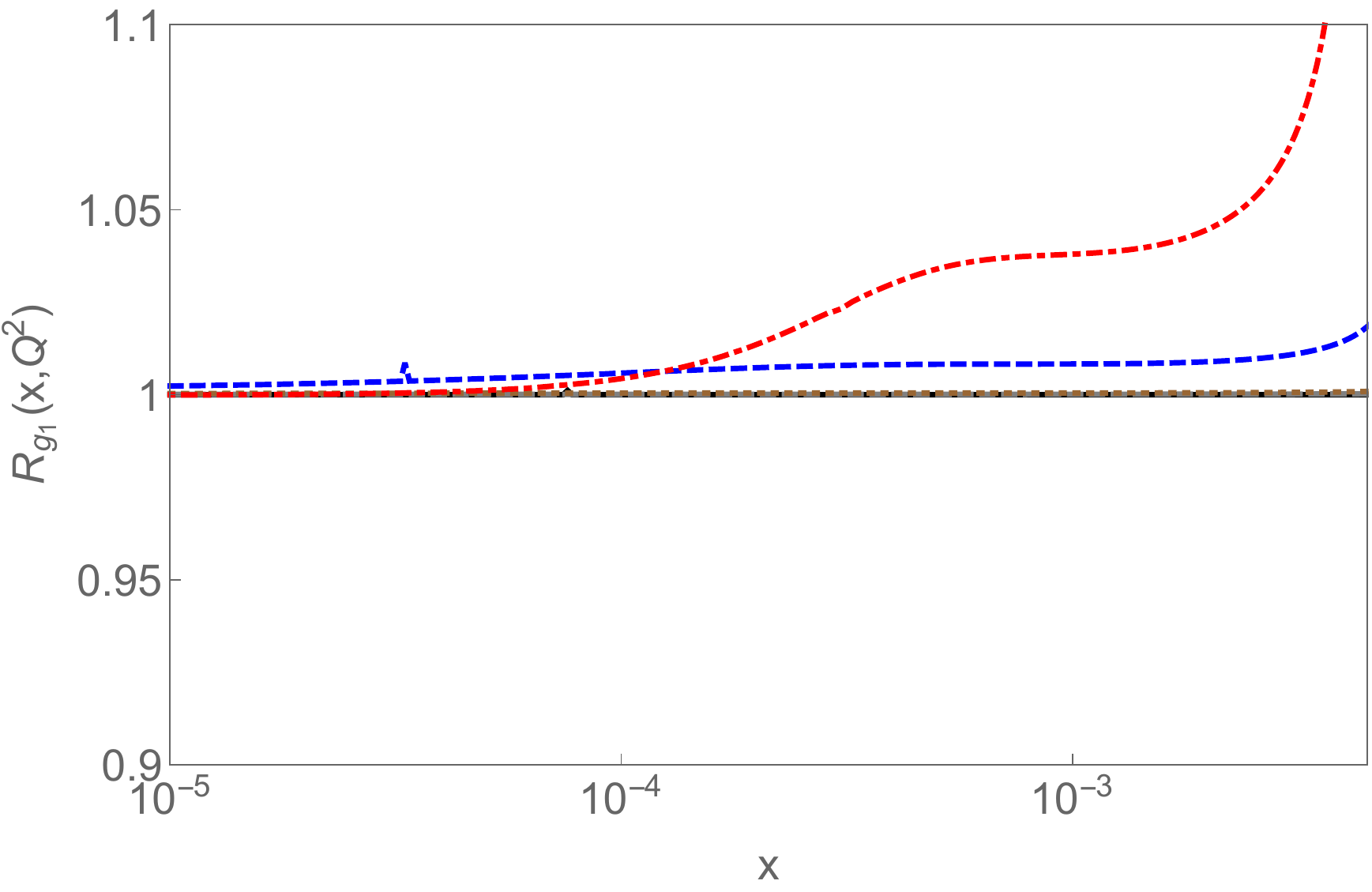}
\includegraphics[width=0.49 \linewidth]{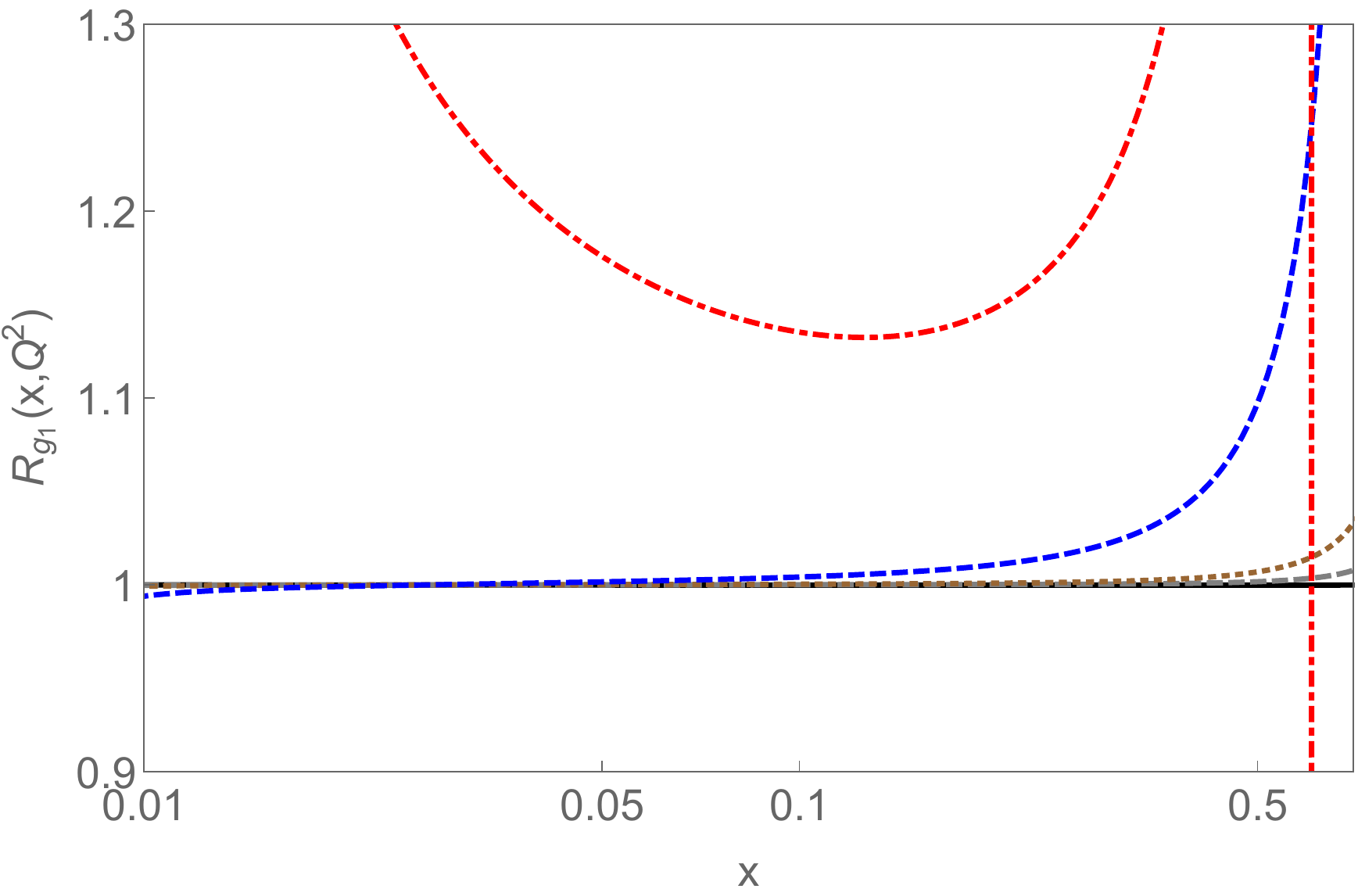}
\caption{\small \sf The ratio of the pure singlet structure function $g_1^{(2),\rm PS}$ 
in the full calculation over the $O((m^2/Q^2)^2)$ improved approximation for different values of $Q^2$ and the 
scale choice $\mu^2 = \mu_F^2 = Q^2$. 
Full lines (Black):         $Q^2=10^4~\GeV^2$; 
dashed lines (Gray):        $Q^2=10^3~\GeV^2$; 
dotted lines (Brown):       $Q^2=500~\GeV^2$; 
dashed lines (Blue):  $Q^2=100~\GeV^2$;
dash-dotted lines (Red):    $Q^2=50~\GeV^2$,
using the parameterization of the parton distribution \cite{Blumlein:2010rn}.}
\label{fig:FIG6}
\end{figure}

In Figure~\ref{fig:FIG6} we depict the ratio of the full result over the $O((m^2/Q^2)^2)$ improved asymptotic
results for $g_1^{\rm PS}$ as a function of $x$ for a series of $Q^2$-values, again separating the small $x$ and 
the large $x$ ranges because of zero transitions for this ratio. 
For $Q^2 \gsim 100 \GeV^2$ the ratios are rather flat and are close to one. The line for $Q^2 = 100 
\GeV^2$ for $x > 0.5$ deviates from one by more than 5\%. Larger deviations are found for $Q^2 = 50 
\GeV^2$, where the 5\% marging is only met for $x < 3 \cdot 10^{-3}$.
We limited the expansion to terms of $\sim O((m^2/Q^2)^2)$ in the present paper, but higher 
order terms can
be given straigtforwardly. The expanded expressions do also allow direct Mellin transforms and provide 
a suitable analytic basis for Mellin--space programmes.

\section{Conclusions}
\label{sec:7}

\vspace*{1mm}
\noindent
We have calculated the massless and massive polarized two--loop pure singlet Wilson coefficients for 
deep-inelastic scattering in analytic form. The calculation has been performed in the Larin--scheme, with a 
final 
finite renormalization to the {\sf M}--scheme, which has been introduced in 
Refs.~\cite{Matiounine:1998re,Moch:2014sna}. The massless Wilson coefficient can be expressed by the 
harmonic 
polylogarithms in $z$--space and harmonic sums in Mellin--$N$ space. In the massive case the polarized 
two--loop pure singlet Wilson coefficient is also given by iterative integrals, however, of a more general kind, 
the Kummer--elliptic integrals, here based on an alphabet of 12 letters, cf.~\cite{Blumlein:2019qze}. From the 
expansion of the massive Wilson coefficient in the region $Q^2 \gg m^2$ one obtains the asymptotic result, which 
can be given in terms of a massive OME and the massless Wilson coefficient, cf.~\cite{POL18}.
In the region of lower values of $Q^2$ and larger values of $x$, the power corrections to the massive 
two--loop Wilson coefficient are essential. From the available analytic result one can construct
the series in $m^2/Q^2$ analytically. Since the deep--inelastic process is usually considered only for 
virtualities $Q^2 \gsim 5~\GeV^2$, this series gives the proper numerical representation in case of the 
charm-quark corrections retaining a relatively small number of terms. The latter representation has the advantage 
that it can be transformed into Mellin space directly, since the expansion coefficients are given in terms of 
harmonic polylogarithms in $z$--space.

\appendix
\section{Contributing Expressions due to Renormalization}
\label{sec:A1}

\vspace*{1mm}
\noindent
In the following we list the Mellin--convolutions, which occurred in Eq.~(\ref{eq:g1full}). These are
convolutions with leading order splitting functions, referring to the parameter $\kappa = m^2/Q^2$.
\begin{eqnarray}
P_{gq}^{(0)} \otimes h_{g_1}^{(1)} &=& C_F T_F \Biggl\{
-192 (1-z) \beta
+32 (1+2 z) \sqrt{1+4 \kappa } \ln \left(\frac{\sqrt{1+4 \kappa }-\beta}{\sqrt{1+4 \kappa }+\beta}\right)
\nonumber \\ &&
+\biggl[
        -64 (1+z) \ln \big(1+\sqrt{1+4 \kappa}\big)
        +64 (1+z) \ln \big(\beta+\sqrt{1+4 \kappa }\big)
\nonumber \\ &&
        -32 (1+z) \ln \left(\frac{\sqrt{1+4 \kappa}-1}{\sqrt{1+4 \kappa}+1}\right)
        +32 (1+z) \ln \left(\frac{\sqrt{1+4 \kappa}-\beta}{\sqrt{1+4 \kappa}+\beta}\right)
\nonumber \\ &&
        -16 \bigl(7-z (1-4 \kappa )\bigr)+32 (1+z) \ln (2)-32 (1+z) \ln (1+\beta )
\biggr] \ln \left(\frac{1-\beta }{1+\beta }\right)
\nonumber \\ &&
-32 (1+z) \text{Li}_2\left(\frac{1-\beta }{2}\right)
+32 (1+z) \text{Li}_2\left(\frac{1+\beta }{2}\right)
-16 (1+z) \ln ^2\left(\frac{1-\beta }{1+\beta }\right)
\nonumber \\ &&
-32 (1+z) \text{Li}_2\left(\frac{1+\beta }{1-\sqrt{1+4 \kappa}}\right)
+32 (1+z) \text{Li}_2\left(\frac{\beta - 1}{\sqrt{1+4 \kappa} - 1}\right)
\nonumber \\ &&
+32 (1+z) \text{Li}_2\left(\frac{1-\beta }{1+\sqrt{1+4 \kappa}}\right)
-32 (1+z) \text{Li}_2\left(\frac{1+\beta }{1+\sqrt{1+4 \kappa}}\right)
\Biggr\}
~,
\\
P_{gq}^{(0)} \otimes \bar{b}_{g_1}^{(1)} &=&
C_F T_F \Biggl\{
208 (1-z) \beta
+\frac{16 (1-k^2)}{k} \ln ^2(1-k)
-\frac{4}{k^2} \big(2 k^2 z -7 k^2-z\big)
\biggl\{
        4 \HA_1 \HA_0
\nonumber \\ &&
      + 2 \ln (1-k) \bigl[ \HA_1 + \HA_{-1} \bigr]
      - 4 \ln (k) \bigl[ \HA_1 + \HA_{-1} \bigr]
      + 2 \ln (1+k) \bigl[ \HA_1 + \HA_{-1} \bigr]
      + \HA_1^2
\nonumber \\ &&
      - 2 \HA_1 \HA_{-1}
      - \HA_{-1}^2
      - 4 \HA_{0,1}
      + 4 \HA_{-1,0}
      + 4 \HA_{-1,1}
\biggr\}
-\frac{8}{k^2} \big(4 k^2+z+7 k^2 z-12 k^2 \beta
\nonumber \\ &&
+12 k^2 z \beta\big) \HA_{-1}
-\frac{8}{k^2} \big(4 k^2+z+7 k^2 z+12 k^2 \beta-12 k^2 z \beta\big) \HA_1
\nonumber \\ &&
+32 (1+2 z)
\biggl\{
      \bigl( 1 + \ln (k) \bigr) \bigl[ \HA_{w_{1}} + \HA_{w_{2}} \bigr]
      - \HA_{w_{1},0}
      - \HA_{w_{2},0}
      - \frac{1}{2} \bigl[ \HA_{w_{1},1} - \HA_{w_{1},-1} 
\nonumber \\ &&
      + \HA_{w_{2},1} - \HA_{w_{2},-1} \bigr]
\biggr\}
- 96 (1-z) \beta \bigl[ \ln (1-k^2) - 2 \ln (k) + 2 \HA_0 \bigr]
\nonumber \\ &&
- 16 \biggl(
         \big(k^2+2 z\big) \ln (1-k)
        +\big(2-k^2+2 z\big) \ln (1+k)
\biggr) \HA_{w_{1}}
\nonumber \\ &&
-16 \biggl(
         \big(2-k^2+2 z\big) \ln (1-k)
        +\big(k^2+2 z\big) \ln (1+k)
\biggr) \HA_{w_{2}}
\nonumber \\ &&
+ 8 (1+z) 
\biggl[
         \HA_1 \HA_{-1}^2
      - 4 \HA_{0,1,1}
      + 4 \HA_{-1,0,1}
      + 8 \HA_{-1,1,0}
      + 8 \HA_{-1,1,1}
      + 4 \HA_{-1,-1,0}
\nonumber \\ &&
      + 8 \HA_{-1,-1,1}
      - \frac{1}{3} \bigl[ \HA_1^3 +  \HA_{-1}^3 \bigr]
      + \bigl[
            \HA_1^2
            - 4 \HA_{-1,1}
      \bigr] \HA_{-1}
      + \bigl[
        4 \HA_{0,1}
      - 4 \HA_{-1,0}
      - 4 \HA_{-1,1}
\nonumber \\ &&
      - 2 \HA_1 \HA_0
      \bigr] \HA_1
      + \bigl[
              \HA_{-1}^2
            - \HA_1^2
            - 2 \HA_1 \HA_{-1}
            + 4 \HA_{-1,1}
      \bigr] \bigl( \ln(1-k^2) - 2 \ln(k) \bigr)
\nonumber \\ &&
      + 2 \bigl( \zeta_2 - \ln^2(2) \bigr) \bigl[ \ln(1-k^2) - \ln(1-z) \bigr]
\biggr]
+ 32 \bigl[ \ln(1-k^2) - \ln(1-z) \bigr] \ln(2)
\nonumber \\ &&
+ 16 k (1+z)
\biggl[
        2 \HA_{w_{1},1,0}
      + \HA_{w_{1},1,1}
      - \HA_{w_{1},1,-1}
      + 2 \HA_{w_{1},-1,0}
      + \HA_{w_{1},-1,1}
      - \HA_{w_{1},-1,-1}
\nonumber \\ &&
      - 2 \HA_{w_{2},1,0}
      - \HA_{w_{2},1,1}
      + \HA_{w_{2},1,-1}
      - 2 \HA_{w_{2},-1,0}
      - \HA_{w_{2},-1,1}
      + \HA_{w_{2},-1,-1}
\nonumber \\ &&
      + \bigl( \ln(1-k^2) - 2 \ln(k) \bigr) \bigl[ \HA_{w_{1},1} + \HA_{w_{1},-1} - \HA_{w_{2},1} - \HA_{w_{2},-1} \bigr]
\nonumber \\ &&
      + \bigl( \zeta_2 - \ln^2(2) \bigr) \bigl[ \HA_{w_{1}} - \HA_{w_{2}} \bigr]
\biggr]
+ 32 k \ln(2) \bigl[ \HA_{w_{1}} - \HA_{w_{2}} \bigr]
\nonumber \\ &&
-\frac{16 (1-k^2) }{k} \biggl( \ln ^2(1+k) +  \ln (1-z) \bigl[ \ln (1-k) - \ln (1+k) \bigr] \biggr)
\Biggr\}
~. 
\end{eqnarray}
Here, $\Li_2(x)$, denotes the dilogarithm \cite{LEWIN},
\begin{eqnarray}
\Li_2(x) = - \int_0^x \frac{dz}{z} \ln(1-z).
\end{eqnarray}

\vspace*{3mm}
\noindent
{\bf Acknowledgment.}\\
We thank would like to thank J.~Ablinger, A.~De Freitas and C.~Schneider for discussions. 
The Feynman diagrams were drawn using {\tt Axodraw} \cite{Collins:2016aya}. This work has been funded in 
part by EU TMR network SAGEX agreement No. 764850 (Marie Sk\l{}odowska-Curie) and COST action CA16201: 
Unraveling new physics at the LHC through the precision frontier, and from the Austrian FWF grants P 27229 
and P 31952.


\end{document}